%% file: main.tex
\documentclass[lettersize,journal]{IEEEtran}
\usepackage{amsmath,amsfonts}
\usepackage{array}
\usepackage[caption=false,font=normalsize,labelfont=sf,textfont=sf]{subfig}
\usepackage{textcomp}
\usepackage{stfloats}
\usepackage{url}
\usepackage{verbatim}
\usepackage{graphicx}
\usepackage{cite}
\usepackage{algorithmicx,algorithm}
\input{macros}

\newcommand\red[1]{\textcolor{black}{#1}}
\begin{document}

\title{RBP-DIP: Residual Back Projection with Deep Image Prior for Ill-Posed CT Reconstruction}

\author{Ziyu Shu, Alireza Entezari
\thanks{Manuscript submitted February 1st, 2024. (corresponding author: Ziyu Shu). This work was supported in part by the NSF [grant number CCF-2210866]}
\thanks{Ziyu Shu is with the CISE department, University of Florida, 32603, USA (email: zs919@nyu.edu).}
\thanks{Alireza Entezari is with the CISE department, University of Florida, 32603, USA (email:entezari@ufl.edu).}}

\markboth{Journal of \LaTeX\ Class Files,~Vol.~14, No.~8, August~2021}%
{Shell \MakeLowercase{\textit{et al.}}: A Sample Article Using IEEEtran.cls for IEEE Journals}

\maketitle

\begin{abstract}
The success of deep image prior (DIP) in a number of image processing tasks has motivated their application in image reconstruction problems in computed tomography (CT). In this paper, we introduce a residual back projection technique (RBP) that improves the performance of deep image prior framework in iterative CT reconstruction, especially when the reconstruction problem is highly ill-posed.

The RBP-DIP framework uses an untrained U-net in conjunction with a novel residual back projection connection to minimize the objective function while improving reconstruction accuracy. In each iteration, the weights of the untrained U-net are optimized, and the output of the U-net in the current iteration is used to update the input of the U-net in the next iteration through the proposed RBP connection. The introduction of the RBP connection strengthens the regularization effects of the DIP framework in the context of iterative CT reconstruction leading to improvements in accuracy. Our experiments demonstrate that the RBP-DIP framework offers improvements over other state-of-the-art conventional IR methods, as well as pre-trained and untrained models with similar network structures under multiple conditions. These improvements are particularly significant in the few-view and limited-angle CT reconstructions, where the corresponding inverse problems are highly ill-posed and the training data is limited. Furthermore, RBP-DIP has the potential for further improvement. Most existing IR algorithms, pre-trained models, and enhancements applicable to the original DIP algorithm can also be integrated into the RBP-DIP framework.

\end{abstract}


\section{Introduction}
\label{sec:introduction}
CT reconstruction algorithms are essential in a number of biomedical imaging applications. The traditional filtered back projection (FBP) algorithm is widely used in commercial scanners as it is computationally efficient. However, its image quality is heavily affected by deviations from ideal imaging conditions (e.g., low-dose, few-view, and limited-angle). To improve image quality in these imaging applications, iterative reconstruction (IR) methods leverage mathematical models for acquisition as well as image priors into an optimization process that forms the basis for image reconstruction~\cite{fessler14}. These methods have been widely studied in CT imaging and are considered the gold standard, their performance in practical imaging applications, robustness, and image quality issues are well understood~\cite{willemink2019evolution}.

Neural network related methods have provided a new set of approaches in CT reconstruction. These methods can be used to improve image quality before (pre-processing, such as sinogram synthesis~\cite{lee2018deep}), during (end-to-end reconstruction~\cite{zhu2018image}, plug in~\cite{shen2019low,shah2018solving}), or after (post processing, such as artifacts removal~\cite{zhang2016image}) reconstruction. The successful application of neural network methods in CT reconstruction is contingent upon access to high-quality datasets and their ability to learn priors that can generalize to instances outside the training dataset. While deep learning methods have had a significant impact in CT imaging research~\cite{wang2018image}, their application in reconstruction poses significant challenges as their lack of stability can lead to the introduction of false structures (hallucinations) that may lead to detrimental effects in medical imaging~\cite{antun2020instabilities, bhadra2021hallucinations}. 

This paper proposes the RBP-DIP framework that inherits the advantages of neural networks' hierarchical structure and conventional IR methods. RBP-DIP employs an untrained U-net as a deep image prior (DIP), combined with the newly proposed residual back-projection (RBP) connection, allowing for optimizing both the network weights and input in different ways (backpropagation through the neural network and conventional IR update through the RBP connection respectively) during each iteration to match the measurements. This provides the RBP-DIP with the following major advantages: 
\begin{itemize}
\item No reliance on training data: the RBP-DIP algorithm employs an untrained neural network, and thus deviates from conventional neural network methodologies. RBP-DIP aligns more closely with IR techniques as it directly minimizes the loss function on inference data. Consequently, the RBP-DIP algorithm needs no training data, therefore circumventing interferences commonly associated with training data and the training process.

\item Combination of DIP and IR: The original DIP algorithm can generate high-quality reconstruction under highly ill-posed conditions. However, it suffers from blurring and neural network-specific artifacts. While IR algorithms suffer from severe artifacts under ill-posed conditions, their stability and interpretability are what DIP lacks. The RBP-DIP algorithm, through the RBP connection, seamlessly combines the strengths of both, achieving superior performance across various conditions.

\item Better reconstruction performance: The experiments show that the RBP-DIP provides significant improvements over conventional IR methods and other neural network models (both untrained and pre-trained) with a similar network structure under multiple conditions. The observed improvements are most significant under few-view and limited-angle conditions.

\item Great potential for further improvement: The proposed framework has great potential for further improvement. It provides a new way of combining the advantages of deep image prior and conventional reconstruction techniques. In this paper, for simplicity, only the basic U-net, residual back projection, and objective function are used for deep image prior, RBP connection, and overall optimization. It is evident that more delicate network structures, IR update techniques, and objective functions can be implemented. Furthermore, the proposed method does not conflict with most existing DIP improvement algorithms. In that case, further improvement is feasible through collaboration with these methods.
\end{itemize}

The remainder of this paper is organized as follows: the details of the CT reconstruction problem and the related works are introduced in Section \ref{sec:related}. The proposed framework is introduced in Section \ref{sec:algorithm}, and compared with conventional methods, pre-trained and untrained methods with a similar network structure on real CT images under various imaging conditions in Section \ref{sec:exp}. The results from our experiments are discussed in Section \ref{sec:diss} followed by the conclusion Section \ref{sec:conclusion}.

\section{Related Works}\label{sec:related}
\subsection{Iterative CT  Reconstruction}\label{sec:CT&MBIR}
IR approaches frame image reconstruction as an optimization problem, minimizing the inconsistency between the forward projection of the iterate and the measurements. X-ray physics, non-ideal effects, and image priors can be incorporated into the reconstruction algorithms by employing various constraints, regularizations, and forward models. The objective function for the CT optimization problem is often expressed by fidelity and regularization terms:
\begin{equation}
	\label{eq:ir}
	\boldsymbol x^* = \argmin_{\boldsymbol x}||\boldsymbol g - \boldsymbol A \boldsymbol x ||^2_2 + \lambda R(\boldsymbol x),
\end{equation}
where $\boldsymbol x$, $\boldsymbol g$, $\boldsymbol A$, and $R$ indicate the image coefficients, sinogram measurements, forward model, and regularization respectively. With accurate forward models~\cite{shu2020gram, shu2022exact, zhang2019convolutional} and effective priors~\cite{sidky2006accurate, jin2010anisotropic, wang2017reweighted, gong2020self, bouman1993generalized, zhang2013gaussian, kisner2012model}, high-quality reconstruction imaging is possible under non-ideal conditions. 

However, when the CT reconstruction is highly ill-posed (e.g., limited-angle and few-view CT reconstruction), the generation of high-quality reconstruction presents difficulties with sparsity-enhancing regularization like total variation, and this is the case even when the measurements are noise-free. In that case, researchers proposed using application-specific priors that are learned from a sample of images to further improve the reconstruction accuracy. Multiple neural network related approaches are developed following this line of thought, and will be discussed in the next section.

\subsection{CT Reconstruction Using Pre-Trained Neural Networks}\label{sec:NN}
Zhu et al.~\cite{zhu2018image} proposed learning a mapping from the acquisition data to the reconstructed images directly. However, the use of fully connected layers makes the computational costs necessary for reconstruction at practical resolutions unacceptable. To circumvent the use of fully connected layers, researchers proposed incorporating additional information into the input to simplify the task so that fully convolutional neural networks are capable of being employed in image reconstruction problems. This is demonstrated by Zhang et al.~\cite{zhang2020pet}, Chen et al~\cite{chen2017low}, and Ye et al.~\cite{ye2018deep}, who suggested utilizing back projections, filtered back projections, and single-view back projections as inputs for high-quality CT reconstruction, respectively.

To further improve the accuracy of reconstructions, researchers proposed designing different neural networks for different kinds of CT reconstruction problems. For ill-posed CT reconstruction, Pan et al.~\cite{pan2024iterative} designed a unique objective function and a convolution-assisted transformer that can simultaneously capture both local and long-range pixel interactions. Wu et al.~\cite{wu2023wavelet} and Zhang et al.~\cite{zhang2024wavelet} proposed improved wavelet denoising techniques to complement the score-based generative model. \red{Li et al. and Wang et al. proposed introducing the diffusion models~\cite{li2024dual,wang2024time}. All these methods have achieved significant results.}

Recently, more sophisticated models have been developed, which involve the collaboration of multiple neural networks working on various subtasks to achieve enhanced performance. Yin et al.~\cite{yin2019domain} introduced an approach that employs two neural networks for denoising in both the sinogram and the image domain. Hu et al.~\cite{hu2021special} and Zhang et al.~\cite{zhang2021clear} extended this concept by incorporating three neural networks. In addition to the two denoising neural networks described in~\cite{yin2019domain}, an auxiliary image domain discriminator for real and reconstructed images was integrated to further refine the model. Similar architectures have also been explored in~\cite{xie2020deep, ma2020low, wu2021drone}.

\subsection{Challenges in Pre-Trained Neural Networks}
Instability has been considered the greatest drawback for neural network image reconstruction. Neural network reconstruction algorithms are designed to learn the prior distribution of the imaged objects from training datasets implicitly or explicitly, so that high-quality reconstructions can be made even with limited and noisy measurements. However, the aforementioned circumstances are predicated upon the assumption that the inference data and training data follow the same distribution. This is particularly problematic for medical image reconstruction as the learned network may omit patient-specific features. Instead, it tends to substitute these unique characteristics with generic features that it has learned from training datasets. Several researches have demonstrated the intrinsic instability of deep learning reconstruction methods, and features can be present in reconstructions that are hallucinated by the network, as investigated by Antun et al.~\cite{antun2020instabilities} and Gottschling et al.~\cite{gottschling2023troublesome}.

Another problem is that current neural network related algorithms are difficult to implement on a large scale. On the one hand, training practical neural networks is challenging. For example, the collaboration of multiple sub-networks utilized in the aforementioned methods necessitates the use of complicated objective functions, which consequently exacerbate the intricacy of the training process. Additionally, to achieve optimal performance, a network must be retrained to learn the best prior for each distinct setting (e.g. reconstruction resolution, detector spacings, noise level). On the other hand, analyzing and improving an existing neural network is also resource-intensive. The absence of a systematic and efficient approach to enhance the performance of a neural network when it fails to meet expectations also presents a challenge. The cascade of multiple sub-networks in the aforementioned methods~\cite{yin2019domain, hu2021special, zhang2021clear, xie2020deep, ma2020low} further complicates these challenges.

\subsection{CT Reconstruction Using Untrained Neural Networks}
To avoid reliance on training data, some researchers proposed methods that require no training data. Ulyanov et al.~\cite{ulyanov2018deep} demonstrated that the hierarchical structure of convolutional networks inherently possesses the capability to capture abundant low-level image statistical priors, thereby enabling the generation of high-quality images. This property is often referred to as the deep image prior (DIP). Several researchers\cite{veen2018compressed,baguer2020computed,shu2022sparse} leveraged the DIP property in their studies on inverse problems. These researchers advocated for minimizing the objective function by optimizing the weights of untrained networks during the reconstruction process. In doing so, the reconstruction is complete once the objective function is minimized. This method was further improved by Shu and Entezari~\cite{shu2022sparse}, where the latent vector $\boldsymbol z$ is updated together with the weights to enhance the convergence speed and reconstruction accuracy. Additionally, they proposed detailed reconstruction instructions and extra regularizations for more accurate reconstructions.

The primary challenge posed by these methods lies in their stability, given that the entire network is initialized randomly and can be trapped in a local minimum. Various strategies have been put forth to overcome this challenge. Venn et al.~\cite{veen2018compressed} proposed running the same algorithm multiple times and choosing the best result; Baguer et al.~\cite{baguer2020computed} recommended the incorporation of supplementary pre-trained neural networks for pre- and post-processing. However, the former approach relies on inefficient trial and error, while the latter reintroduces pre-trained models, contravening its initial objective. 

\begin{figure*}
	\centering
	\includegraphics[width=\linewidth]{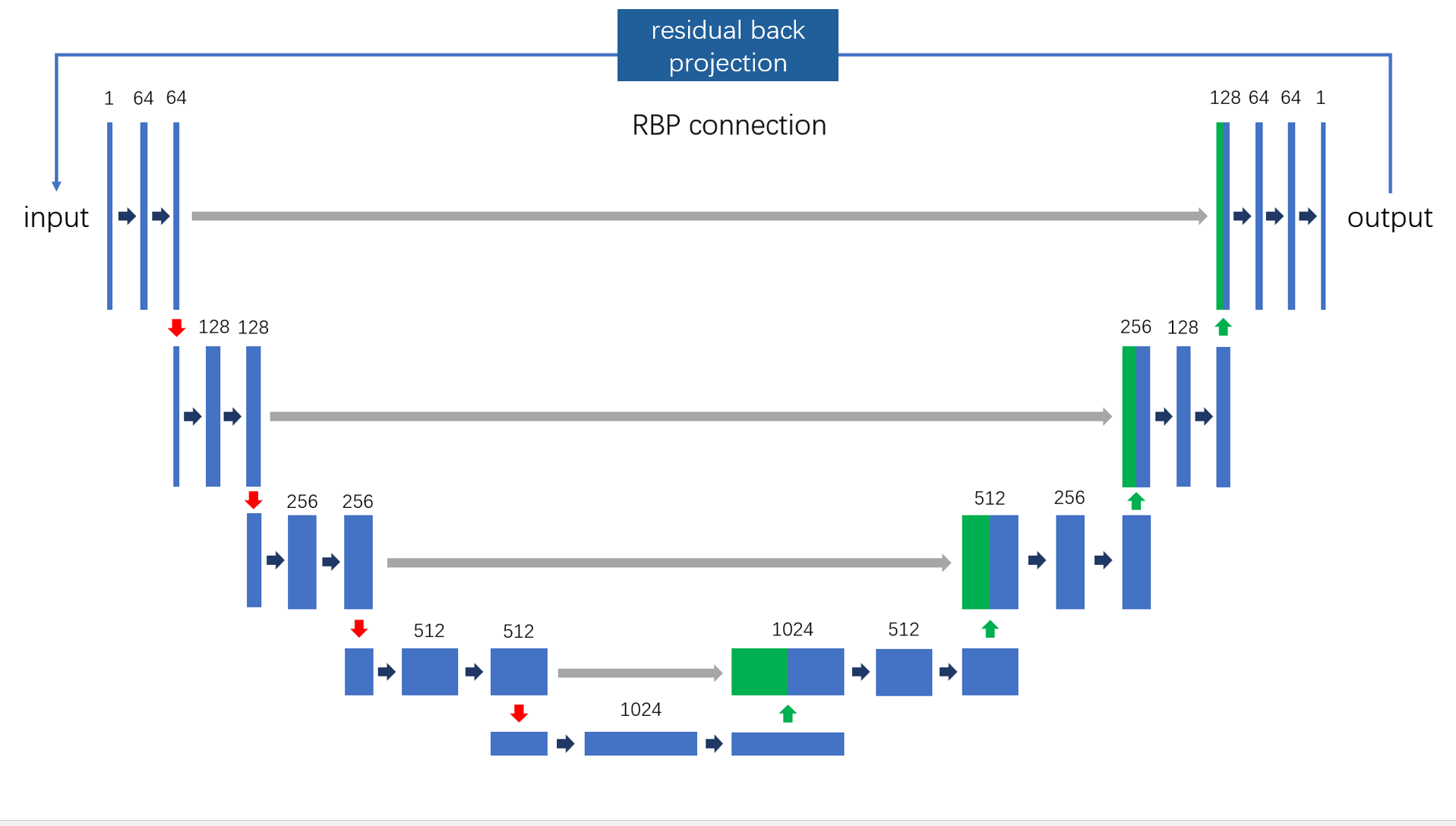}
	\caption{The architecture of the RBP-DIP framework: In each iteration, a conventional IR algorithm can be used to calculate the necessary updates for the current reconstruction based on the neural network's output. This information is then passed to the neural network's input in the next iteration through the RBP connection, indirectly affecting the network's subsequent output. It is worth mentioning that the DIP algorithm, which is a highly nonlinear algorithm based on convolutional neural networks, uses different prior compared to conventional IR algorithms. This difference allows the information transmitted by the RBP connection to be more effectively utilized by the DIP algorithm, significantly enhancing its performance.}
	\label{Unet}
\end{figure*}

\section{Methods}\label{sec:algorithm}
\subsection{CT Reconstruction with Deep Image Prior}
As discussed before, DIP in CT reconstruction optimizes an objective function similar to the conventional IR method~\cite{veen2018compressed}, which can be formulated as:
\begin{equation}
	\begin{aligned}
		\label{eq:csdip}
		\boldsymbol w^* &= \argmin_{\boldsymbol w}||\boldsymbol g - \boldsymbol A G(\boldsymbol w; \boldsymbol z) ||^2_2,\\ \boldsymbol{x}^* &= G(\boldsymbol w^*; \boldsymbol z),
	\end{aligned}
\end{equation}
where $G(\boldsymbol w; \boldsymbol z)$ represents an untrained convolutional neural network, it takes $\boldsymbol z$ as the input and is parameterized by the weights $\boldsymbol w$. Here $\boldsymbol z$ is a Gaussian vector that is randomly initialized and then fixed.

The key difference between the conventional IR method (Equation \ref{eq:ir}) and the DIP method (Equation \ref{eq:csdip}) is that the DIP method uses the output of $G(\boldsymbol w; \boldsymbol z)$ to represent the reconstruction result $\boldsymbol x$. DIP related methods utilize the untrained neural network $G$ to minimize objective function by optimizing the network weights. In that case, the reconstruction result is constrained to lie in the space spanned by that untrained neural network, so that the DIP property mostly ensures the generation of high-quality results, especially when the corresponding inverse problem is highly ill-posed.  

It is worth mentioning that both DIP and IR methods minimize a similar loss function (regularizations can also be applied to Equation \ref{eq:csdip}~\cite{liu2019image}) on inference data directly. This ensures that the results of DIP are not inferior to those of IR under similar loss conditions. However, DIP related methods are susceptible to blurring and neural network specific artifacts. These shortcomings are demonstrated in Fig.\ref{reconresult}d, and Fig.\ref{reconresultfan}d, where the reconstruction results of the DIP method suffer from abnormal artifacts when the corresponding inverse problem is highly ill-posed, and lack image details even when the problem is well-posed (full-view CT reconstruction).

\subsection{Residual Back Projection with Deep Image Prior}
\label{sec:rbpdip}

\begin{algorithm}[t]
	\caption{RBP-DIP}
	\label{alg1}
	\hspace*{0.02in} {\bf Input:} 
	measure matrix $\boldsymbol A$, measurement $\boldsymbol g$, output $\boldsymbol x$ (zero at the beginning), input $\boldsymbol z$ (zero at the beginning), and step size $\beta$. The untrained model $G$ indicates the U-net in Fig.\ref{Unet} without the RBP connection.
	\begin{algorithmic}[1]
		\State {\bf Repeat:}  
		\State \quad $\boldsymbol r = \boldsymbol A^{\rm T} (\boldsymbol g - \boldsymbol A \boldsymbol x)$ \quad \slash * Residual Back projection *\slash
		\State \quad $\boldsymbol z = \boldsymbol z + \beta  \frac{\boldsymbol r}{||\boldsymbol r||_2^2}$ \quad \quad \slash * Network input modification *\slash
        \State \quad $\boldsymbol z = \frac{\boldsymbol z}{||\boldsymbol z||_2^2} $ \quad \, \quad \quad \quad \slash * Network input normalization *\slash
        \State \quad $\boldsymbol x = G(\boldsymbol w; \boldsymbol z)$ \quad \quad \quad \slash * U-net forward propagation *\slash
        \State \quad Update $G$, using $||\boldsymbol g - \boldsymbol A \boldsymbol x||_2^2$ as the loss function. 
        
        \quad \quad \quad \quad \, \,  \quad \quad \quad \slash * U-net backpropagation *\slash
        \State {\bf Optional:} apply IR algorithm on $\boldsymbol x$.
	\end{algorithmic}
\end{algorithm}

Since its inception, researchers have been trying to further improve the DIP algorithm by adjusting its network input. Ulyanov et al.~\cite{ulyanov2018deep} introduced small Gaussian noise into the network input in each iteration to improve the algorithm's stability and noise resistance. Shu and Entezari's research~\cite{shu2022sparse} showed that optimizing both the network input and network weights can further improve the convergence speed and reconstruction accuracy of the DIP method. Cui et al.~\cite{cui2022unsupervised}, in conditional DIP (CDIP), proposed using a reference image as the network input to guide the DIP reconstruction process. These studies imply that the network input can significantly impact the DIP algorithm, and using a fixed random vector may not be the best option. However, existing algorithms each have their shortcomings. The improvement provided by the method proposed in~~\cite{shu2022sparse} is somewhat limited, mainly because the algorithm homogenizes the network input and network weights — they are both randomly initialized and updated via the neural network's backpropagation algorithm. Given that the number of parameters for the network input is much smaller than that of the network weights, the impact of updating the network input is limited. CDIP algorithm's effectiveness highly depends on the quality of the reference image. However, high-quality reference images similar to the ground truth image are often difficult to obtain (since the algorithm's goal is to achieve high-quality reconstruction similar to the ground truth image). Furthermore, if the reference image itself contains artifacts, these artifacts will mislead the DIP algorithm. 

\red{Nevertheless, these algorithms have demonstrated that modifying the input of the DIP network can significantly affect the reconstruction process and its result. To further leverage this property while overcoming the challenges faced by the aforementioned algorithms, we propose the RBP-DIP framework. The basic idea of the RBP-DIP is to update the network input based on the network output in the previous iteration. RBP-DIP is similar to the method proposed in~\cite{shu2022sparse}, as both involve updating the network input and network weights. The difference is that RBP-DIP uses different update methods for the network input and network weights: network weights are updated through the highly nonlinear neural network by the backpropagation algorithm, while the network input is updated using a mostly linear IR algorithm. This addresses the homogeneity issue of network weight and network input in DIP methods. Furthermore, since different update algorithms are used, the information contained in the network input is unobtainable by the original DIP algorithm. Hence, the updated network input can also be regarded as a reference image to guide the reconstruction of the DIP algorithm. This also aligns with CDIP, but differs in that the reference image is automatically generated by RBP-DIP during reconstruction, eliminating the need for a high-quality reference image and avoiding potential interference from a suboptimal reference image.}

The architecture of our framework, residual back projection with the deep image prior (RBP-DIP), is shown in Fig.\ref{Unet}. It contains a U-net and an RBP connection. The U-net, which has demonstrated its efficacy in various image processing problems, is used to implement the standard DIP algorithm in our approach. In each iteration, the input of the U-net is iteratively updated using the back projected residual between the sinogram of the current network output and the provided measurement via the RBP connection. This update incorporates the information from the conventional IR algorithm and can be used to enhance the output of the U-net. The RBP connection constitutes the primary distinction between our RBP-DIP model and the DIP model presented in~\cite{veen2018compressed,shu2022sparse}. Notably, the RBP connection does not participate in the backpropagation process. Also, batch normalization is still implemented in the RBP-DIP network. Although RBP-DIP directly optimizes the inference data, making its batch size equal to $1$, batch normalization can still effectively address the issues of gradient explosion and gradient vanishing, making the neural network easier to converge.

An intuitive explanation of RBP-DIP is that it uses a conventional IR method to update network input $\boldsymbol z$, and uses the network output $\boldsymbol x$ as the input of the conventional IR method. Such an architecture is designed to take advantage of both the conventional IR method and the DIP method. The DIP property can help the untrained U-net generate high-quality images that are free from most artifacts existing in conventional IR methods, especially when the corresponding inverse problem is highly ill-posed. However, the DIP method is always considered unstable and may generate neural network-specific artifacts, as the whole neural network is randomly initialized and cannot be clearly explained. On the contrary, although the conventional IR methods are unable to generate accurate reconstruction results when the inverse problem is highly ill-posed, its convergence rate, stability, and robustness have already been well-analyzed. Thus, it can be used to guide the DIP method, correct the aforementioned network-specific artifacts, fine-tune the reconstruction result, and increase the robustness of the whole framework. As a result, the integration of the RBP connection and DIP method can amalgamate the advantages of both methods and mitigate each other's shortcomings.

\begin{figure*}
	\centering
	\begin{minipage}[b]{.39\linewidth}
		\centering
		\centerline{\includegraphics[width=\linewidth]{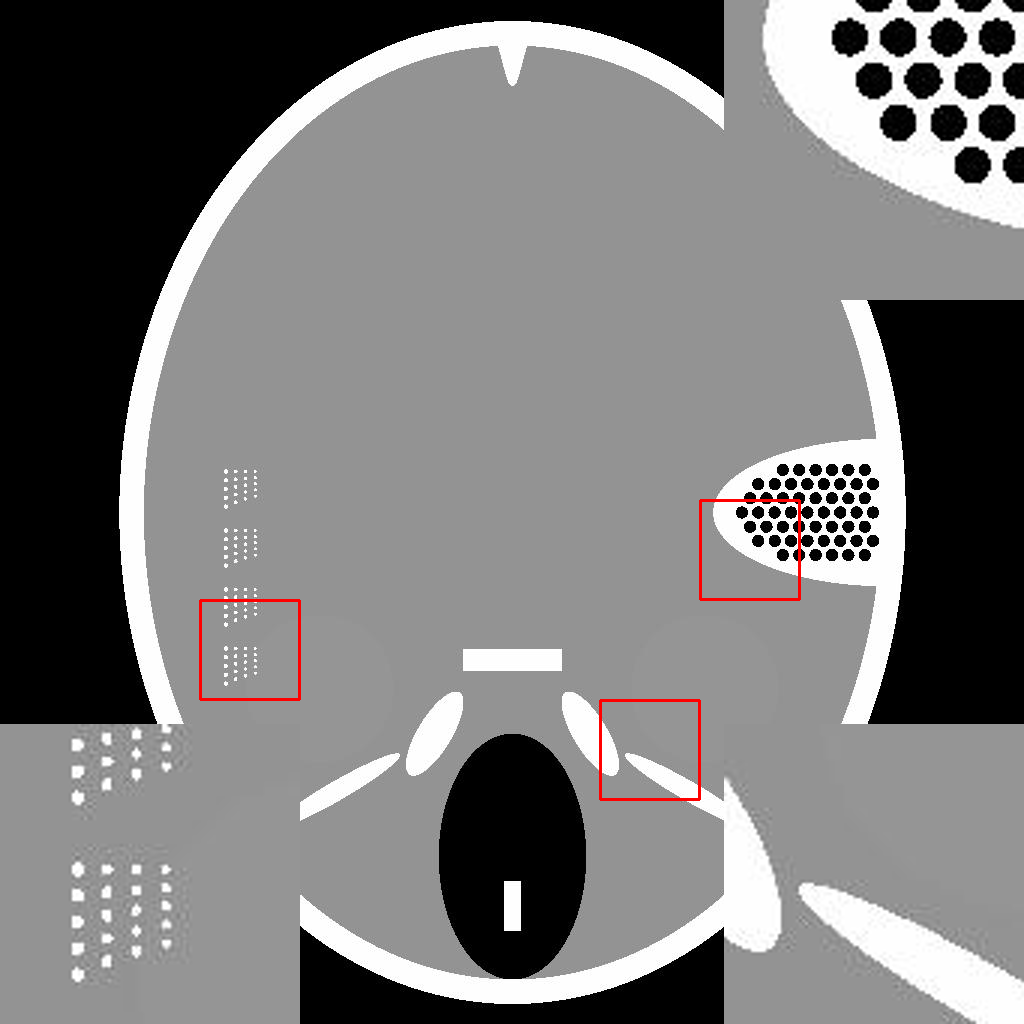}}
		  \vspace{0.5cm}
		\centerline{(a) Ground Truth}\medskip
	\end{minipage}
	\begin{minipage}[b]{.19\linewidth}
		\begin{minipage}[b]{1\linewidth}
		\centering
		\centerline{\includegraphics[width=\linewidth]{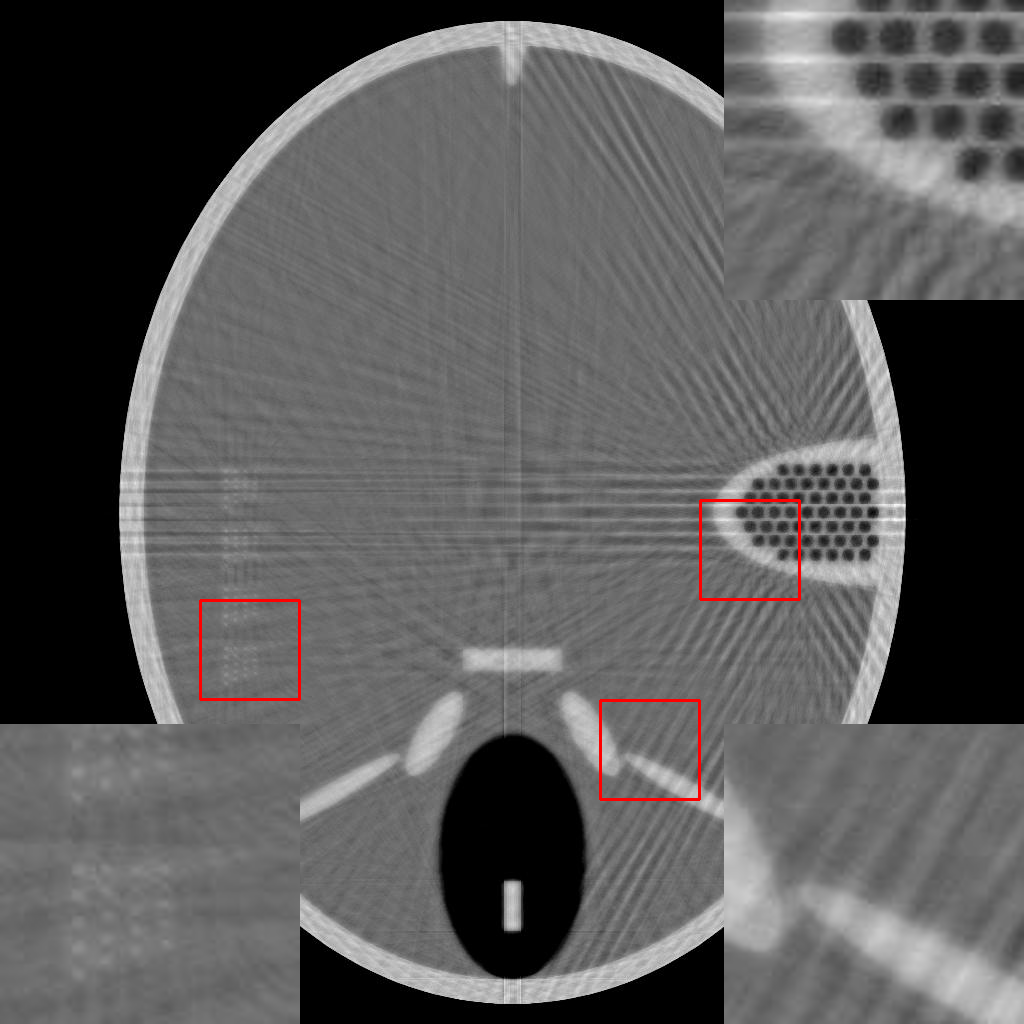}}
		\centerline{(b) 19.01dB}\medskip
	\end{minipage}
 
    \begin{minipage}[b]{1\linewidth}
		\centering
		\centerline{\includegraphics[width=\linewidth]{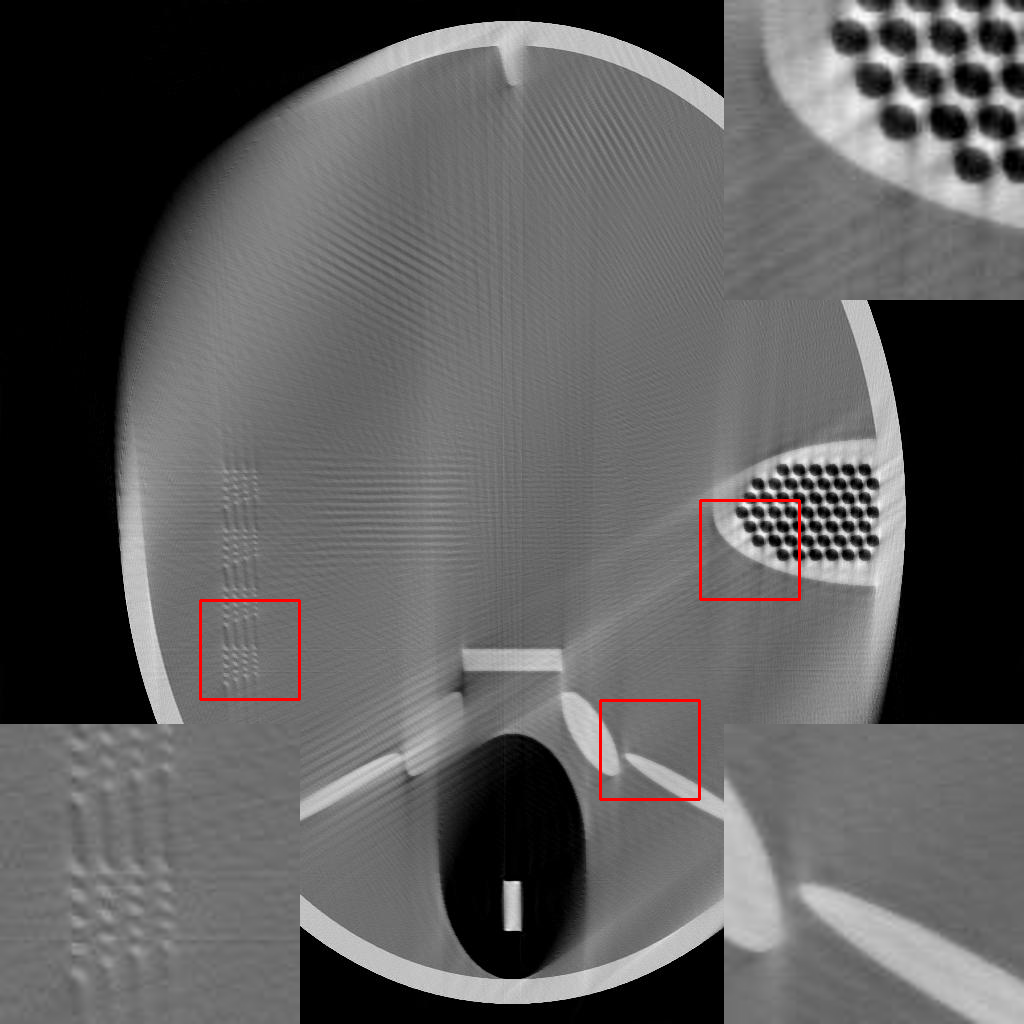}}
		\centerline{(e) 15.35dB}\medskip
	\end{minipage}
	\end{minipage}
	\begin{minipage}[b]{.19\linewidth}
		\begin{minipage}[b]{1\linewidth}
		\centering
		\centerline{\includegraphics[width=\linewidth]{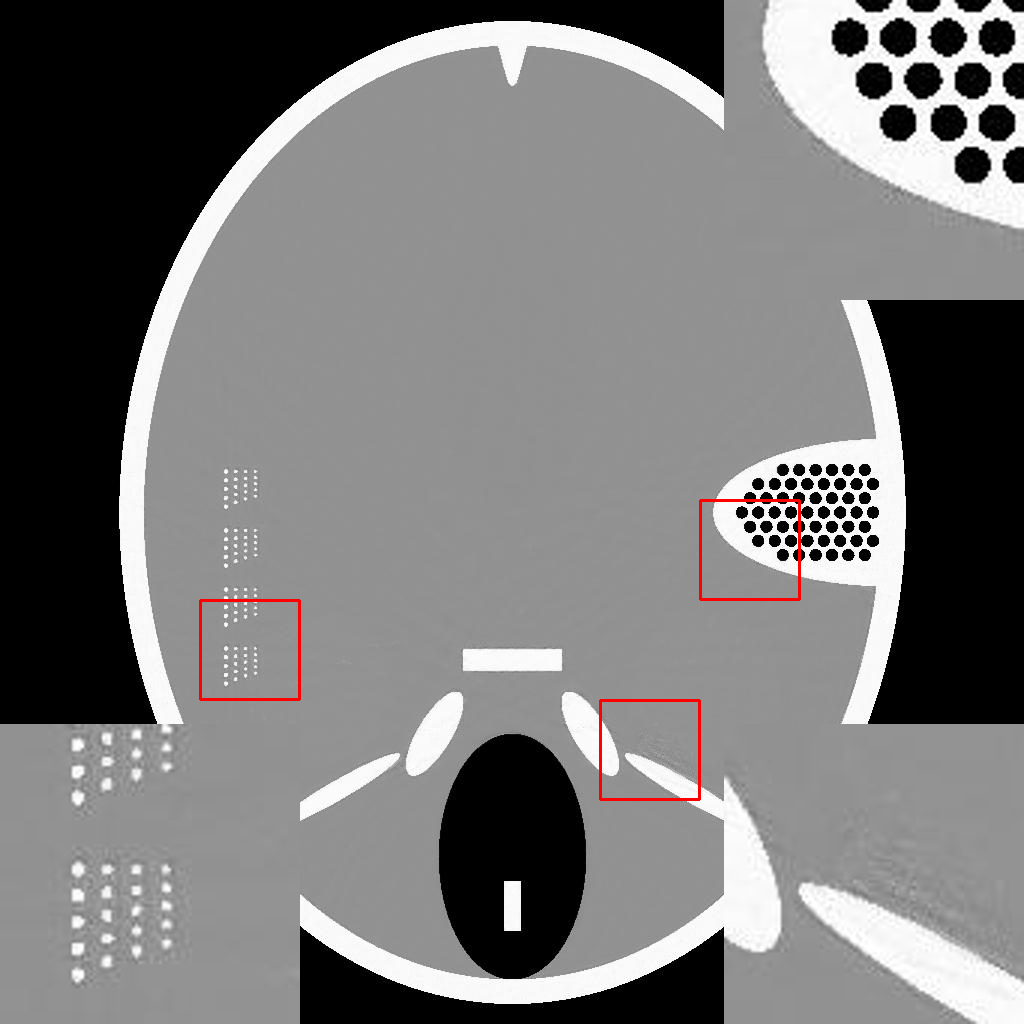}}
		\centerline{(c) 37.72dB}\medskip
	\end{minipage}
 
    \begin{minipage}[b]{1\linewidth}
		\centering
		\centerline{\includegraphics[width=\linewidth]{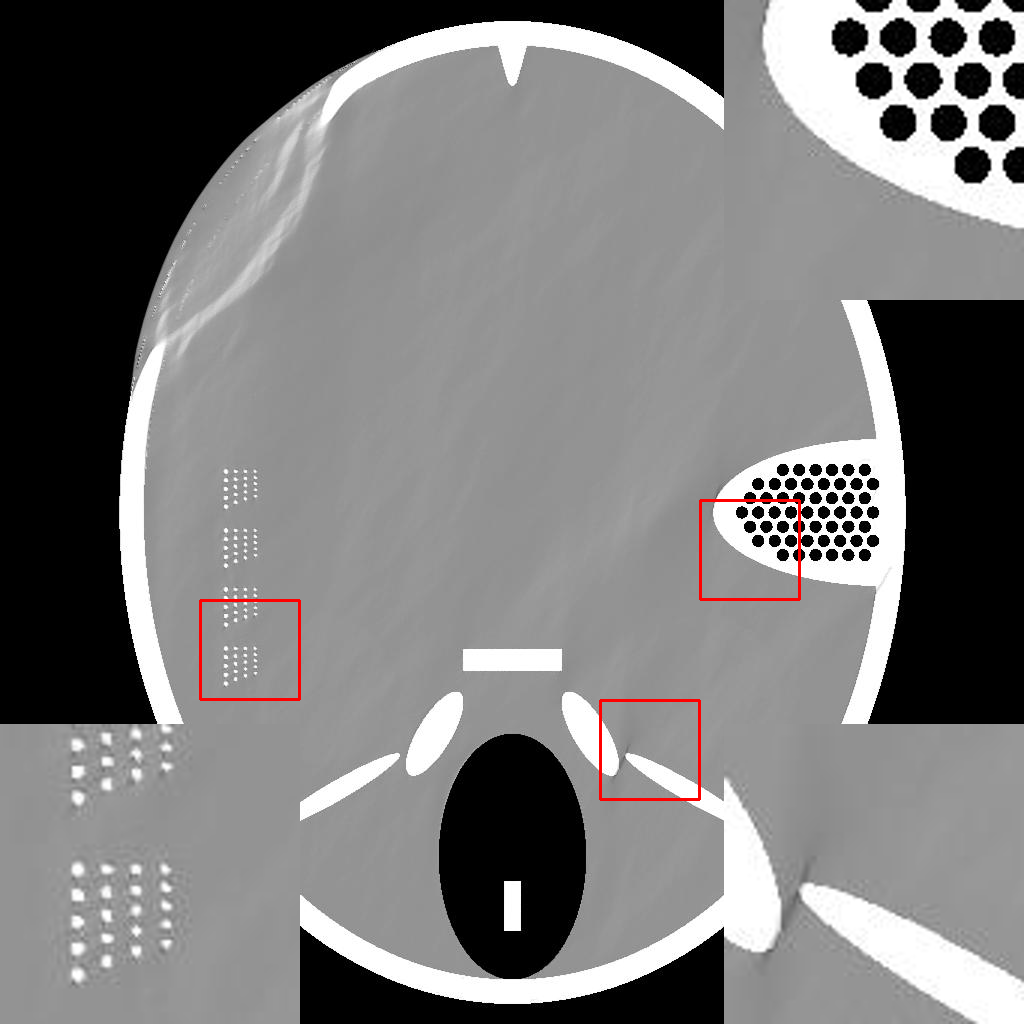}}
		\centerline{(f) 20.54dB}\medskip
	\end{minipage}
	\end{minipage}
	\begin{minipage}[b]{.19\linewidth}
		\begin{minipage}[b]{1\linewidth}
		\centering
		\centerline{\includegraphics[width=\linewidth]{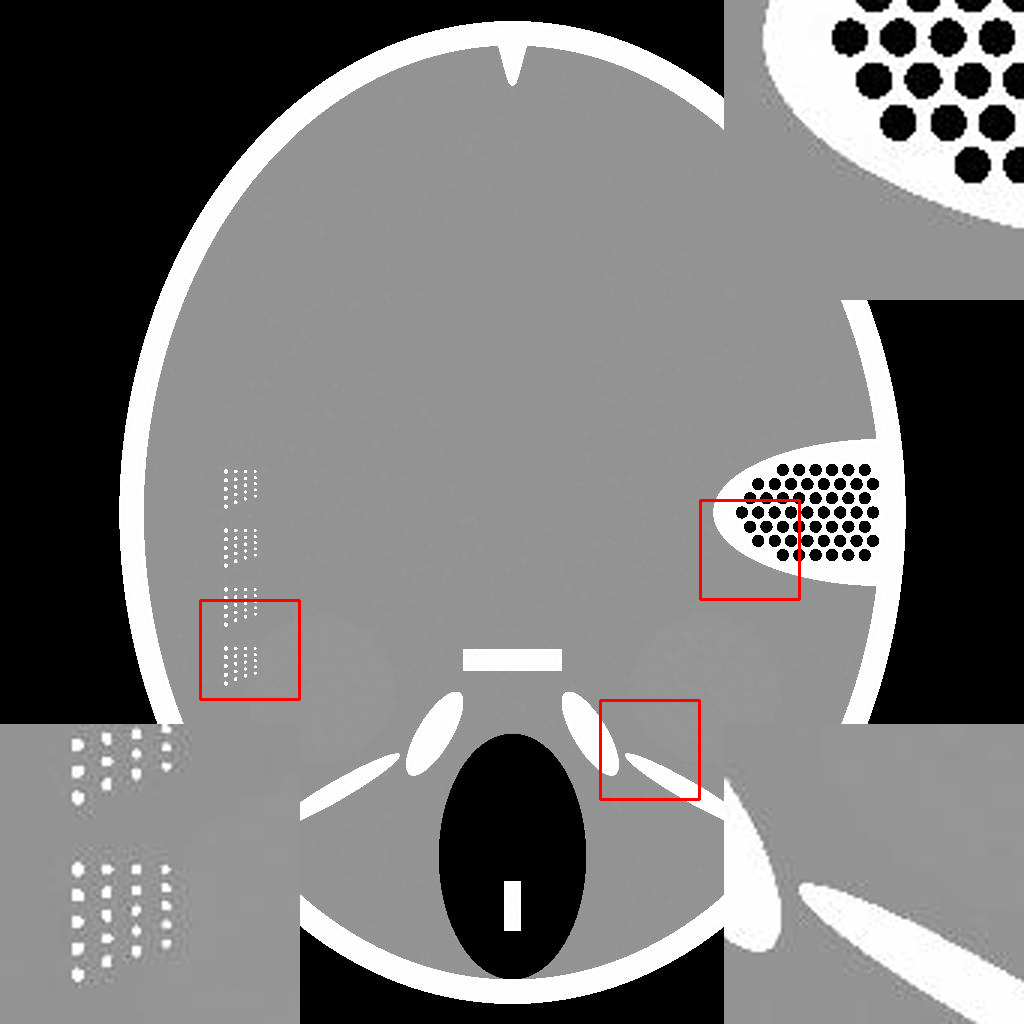}}
		\centerline{(d) 38.96dB}\medskip
	\end{minipage}
 
    \begin{minipage}[b]{1\linewidth}
		\centering
		\centerline{\includegraphics[width=\linewidth]{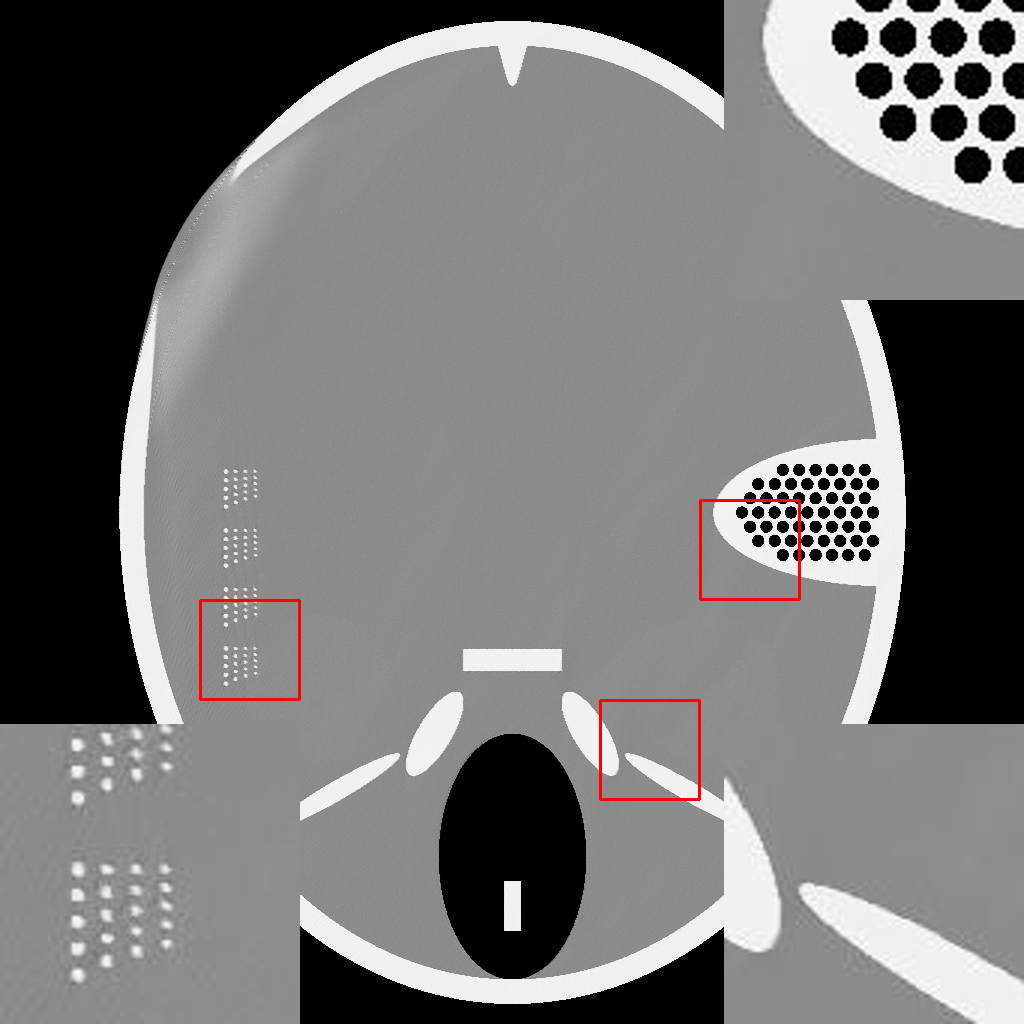}}
		\centerline{(g) 22.84dB}\medskip
	\end{minipage}
	\end{minipage}

	\caption{The few-view (first row, $60$ views which are evenly distributed from $0^\circ$ to $360^\circ$) and limited-angle (second row, $120$ views which are evenly distributed from $0^\circ$ to $120^\circ$) reconstruction results of the Forbild phantom for ASD-POCS (b and e), DIP (c and f), and RBP-DIP (d and g). The proposed RBP-DIP method can generate high-quality reconstruction results with minimal artifacts.}
	\label{Fresult}
\end{figure*}

The details of RBP-DIP are shown in Algorithm \ref{alg1}, where lines 2-4 correspond to the RBP connection. $\boldsymbol r$ in line 2 indicates the back-projected residual, which is the derivative of the data inconsistency $ || \boldsymbol g - \boldsymbol A \boldsymbol x ||^2$ with respect to $\boldsymbol x$ and is also a major component of most IR algorithms' updating equation. In the RBP-DIP algorithm, instead of updating $\boldsymbol x$ directly, $\boldsymbol r$ is used by the RBP connection to update the network's input $\boldsymbol z$. \red{The reason is that directly updating $\boldsymbol x$ will result in streak artifacts, as shown in Fig.\ref{Fresult}. Updating $\boldsymbol z$ allows the deep image prior to help eliminate these artifacts. Additionally, it is postulated that the iteratively updated vector $\boldsymbol z$ can more effectively guide the model since it contains the information provided by the IR algorithm embedded in the RBP connection.} Lines 5 and 6 correspond to the optimization of the untrained U-net using the backpropagation algorithm.

This paper focuses on demonstrating the advantages of the proposed RBP connection and the corresponding RBP-DIP framework over the original DIP algorithm. It also shows the great potential of replacing the original DIP algorithm with the RBP-DIP algorithm in most DIP-related algorithms. Therefore, we propose using the basic setup similar to the original DIP algorithm. Thus, the objective function of RBP-DIP is just the data inconsistency $||\boldsymbol g - \boldsymbol A \boldsymbol x ||^2_2$, and the IR algorithm integrated into the RBP connection is a simple back projection $\boldsymbol r = \boldsymbol A^{\rm T} (\boldsymbol g - \boldsymbol A \boldsymbol x)$. We believe that if the RBP-DIP framework can succeed in the simplest setup, then it can achieve even better improvements with more sophisticated objective functions, IR algorithms (in the RBP connection), and network structures.

It is worth mentioning that substantial fluctuations in the input to the U-net could undermine its stability and overall performance. Consequently, we normalize the input $\boldsymbol z$ in line 4 to let the network input always be a unit vector. Moreover, the magnitude of RBP updates necessitates regulation, as the RBP updates would otherwise dominate the reconstruction process and lead to artifacts commonly seen in conventional IR methods. In that case, the residual $\boldsymbol r$ is also normalized in lines 3. Together with the function $\beta$, the RBP update in line 3 is a summation of the previous unit vector $z$ and another vector of length $\beta$, which can be expressed as:
\begin{equation}
	\beta=\frac{10^{-3}}{1+e^{-(\frac{n}{n_s}-n_c)}},
\end{equation}
where $n$ indicates the current number of iterations. $\beta$ is a sigmoid function centered at $n_c$ and stretched by the factor $n_s$. In the initial phase of the reconstruction procedure ($\frac{n}{n_s} \ll n_c$), the DIP property plays a crucial role in generating high-quality preliminary reconstruction outcomes. During this period, the RBP connection is suppressed by the factor $\beta$ to avoid introducing artifacts commonly associated with IR methods. As the reconstruction progresses and $\frac{n}{n_s} \gg n_c$, the RBP connection's strength is augmented by the increasing $\beta$. In this later stage, the RBP connection utilized its embedded IR method to further refine the reconstruction without engendering additional artifacts, given that a high-quality preliminary reconstruction has already been generated. To achieve optimal performance, it is imperative for $n_c$ to be sufficiently large to yield a smooth function, preventing significant disruptions to the network stability caused by input updates. The selection of $n_c$ mainly depends on the nature of the inverse problem itself. In instances where the inverse problem is highly ill-posed, such as few-view and limited-angle CT reconstruction, conventional iterative reconstruction (IR) methods implemented in the proposed RBP connection may result in pronounced artifacts. In such cases, it is advisable to use a larger $n_c$. This adjustment gives the algorithm more time to leverage the properties of DIP to obtain a better initial guess. Conversely, if the inverse problem is closer to being well-posed or over-determined, a smaller $n_c$ enables the RBP to take in the optimization process more promptly. In this paper, these parameters are approximately determined through our experiments and ensure that all algorithms perform well. While adjusting hyperparameters for different application scenarios can further enhance the performance, the gains from such fine-tuning are insignificant compared to the improvements achieved by RBP-DIP based on DIP. Moreover, this paper primarily aims to propose an RBP-DIP framework based on the original DIP algorithm, rather than optimizing the network for a specific image reconstruction task. As a result, we have set the parameters to $n_s = 250$, $n_c = 10$, and the total number of iterations is $5000$. By employing these values, the RBP connection is suppressed during the first half of the reconstruction process and gradually augmented during the second half.

RBP-DIP, being related to DIP, currently lacks comprehensive theoretical proof of convergence.  However, numerous experiments~\cite{ulyanov2018deep,veen2018compressed,baguer2020computed,shu2022sparse,mataev2019deepred,gong2018pet} by different researchers have also shown that the DIP network converges in most cases. The main difference between the RBP-DIP and the original DIP algorithm lies in the iterative updates to the network input $\boldsymbol z$. Several studies~\cite{ulyanov2018deep, mataev2019deepred, shu2024sdip, cui2022unsupervised } have shown that this not only retains the stability and convergence performance of the algorithm but also improves the algorithm's noise resistance. In fact, in many DIP-related papers, including the original DIP paper, modifications are made by authors to the vector $\boldsymbol z$ in each iteration, such as adding a small Gaussian noise vector, to enhance the robustness of the algorithm. The RBP-DIP algorithm updates $\boldsymbol z$ with more meaningful information to achieve better performance, as demonstrated by our experiments. Furthermore, to mitigate potential adverse effects, we propose a conventional IR algorithm at line 7 as an optional post-processing step. This aims to ensure convergence. However, in all subsequent experiments presented in this paper, we omit the post-processing step and still observed excellent results.

\begin{figure*}
    \centering
    \begin{minipage}[b]{.19\linewidth}
        \centering
        \centerline{\includegraphics[width=\linewidth]{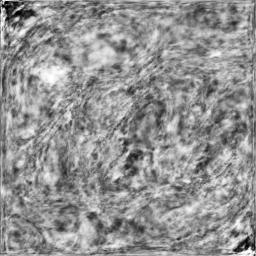}}
    \end{minipage}
    \begin{minipage}[b]{.19\linewidth}
        \centering
        \centerline{\includegraphics[width=\linewidth]{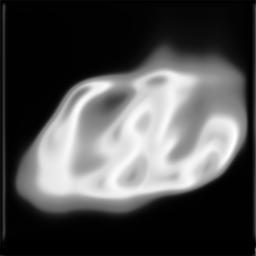}}
    \end{minipage}
    \begin{minipage}[b]{.19\linewidth}
        \centering
        \centerline{\includegraphics[width=\linewidth]{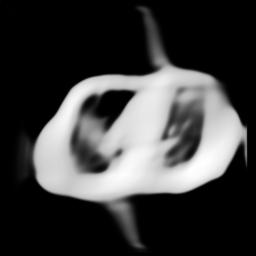}}
    \end{minipage}
    \begin{minipage}[b]{.19\linewidth}
        \centering
        \centerline{\includegraphics[width=\linewidth]{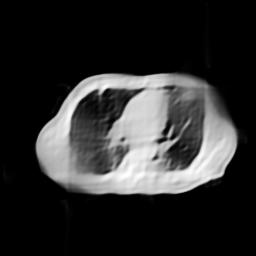}}
    \end{minipage}
    \begin{minipage}[b]{.19\linewidth}
        \centering
        \centerline{\includegraphics[width=\linewidth]{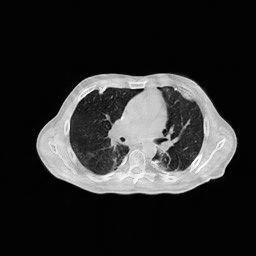}}
    \end{minipage}

    \begin{minipage}[b]{.19\linewidth}
        \centering
        \centerline{\includegraphics[width=\linewidth]{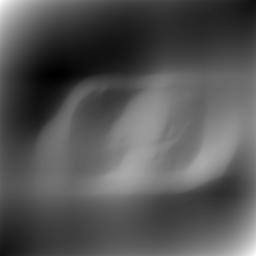}}
        \centerline{(a) 1st Iteration}\medskip
    \end{minipage}
    \begin{minipage}[b]{.19\linewidth}
        \centering
        \centerline{\includegraphics[width=\linewidth]{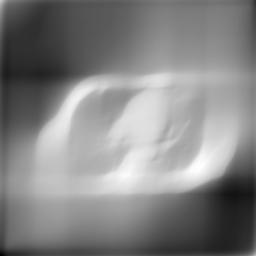}}
        \centerline{(b) 10th Iteration}\medskip
    \end{minipage}
    \begin{minipage}[b]{.19\linewidth}
        \centering
        \centerline{\includegraphics[width=\linewidth]{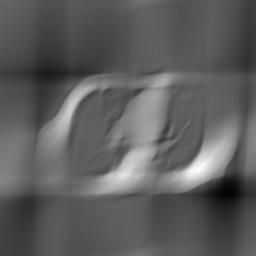}}
        \centerline{(c) 20th Iteration}\medskip
    \end{minipage}
    \begin{minipage}[b]{.19\linewidth}
        \centering
        \centerline{\includegraphics[width=\linewidth]{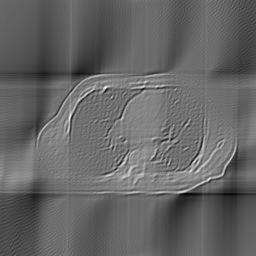}}
        \centerline{(d) 2000th Iteration}\medskip
    \end{minipage}
    \begin{minipage}[b]{.19\linewidth}
        \centering
        \centerline{\includegraphics[width=\linewidth]{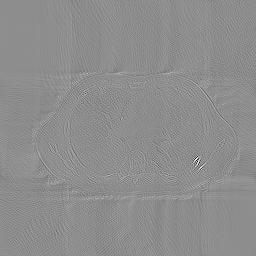}}
        \centerline{(e) 10000th Iteration}\medskip
    \end{minipage}
    
    \caption{The output (first row) and the corresponding input $\boldsymbol z$ (second row) at different iteration. The number of views is $90$, uniformly distributed from $0^{\circ}$ to $90^{\circ}$. Each figure in the second row has its histogram individually adjusted for better visualization.}
    \label{RBP recon process}
\end{figure*}

\begin{figure}[htb]
    \begin{minipage}[b]{\linewidth}
        \centering
        \centerline{\includegraphics[width=\linewidth]{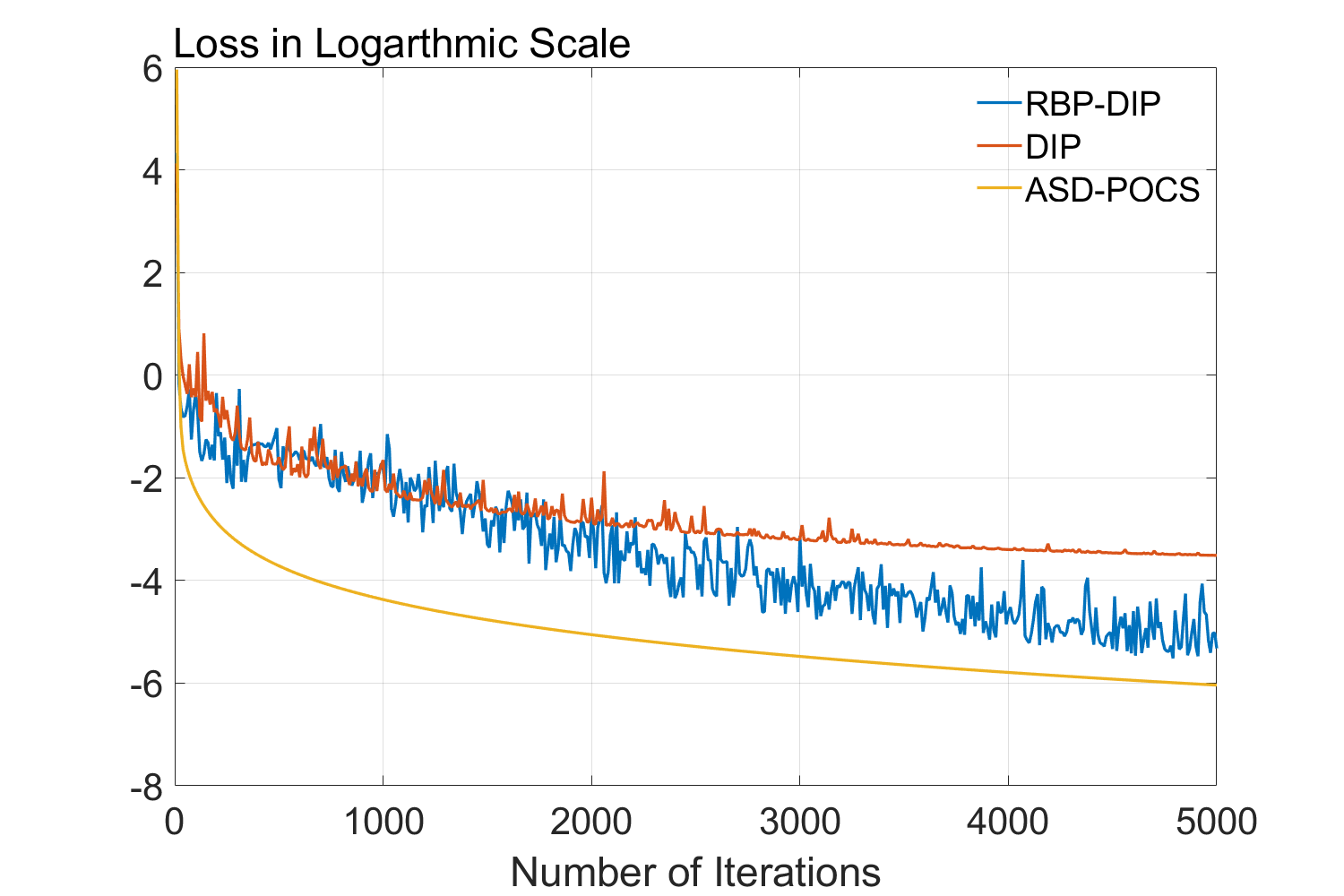}}
        \centerline{(a)}\medskip
    \end{minipage}
    
    \begin{minipage}[b]{\linewidth}
        \centering
        \centerline{\includegraphics[width=\linewidth]{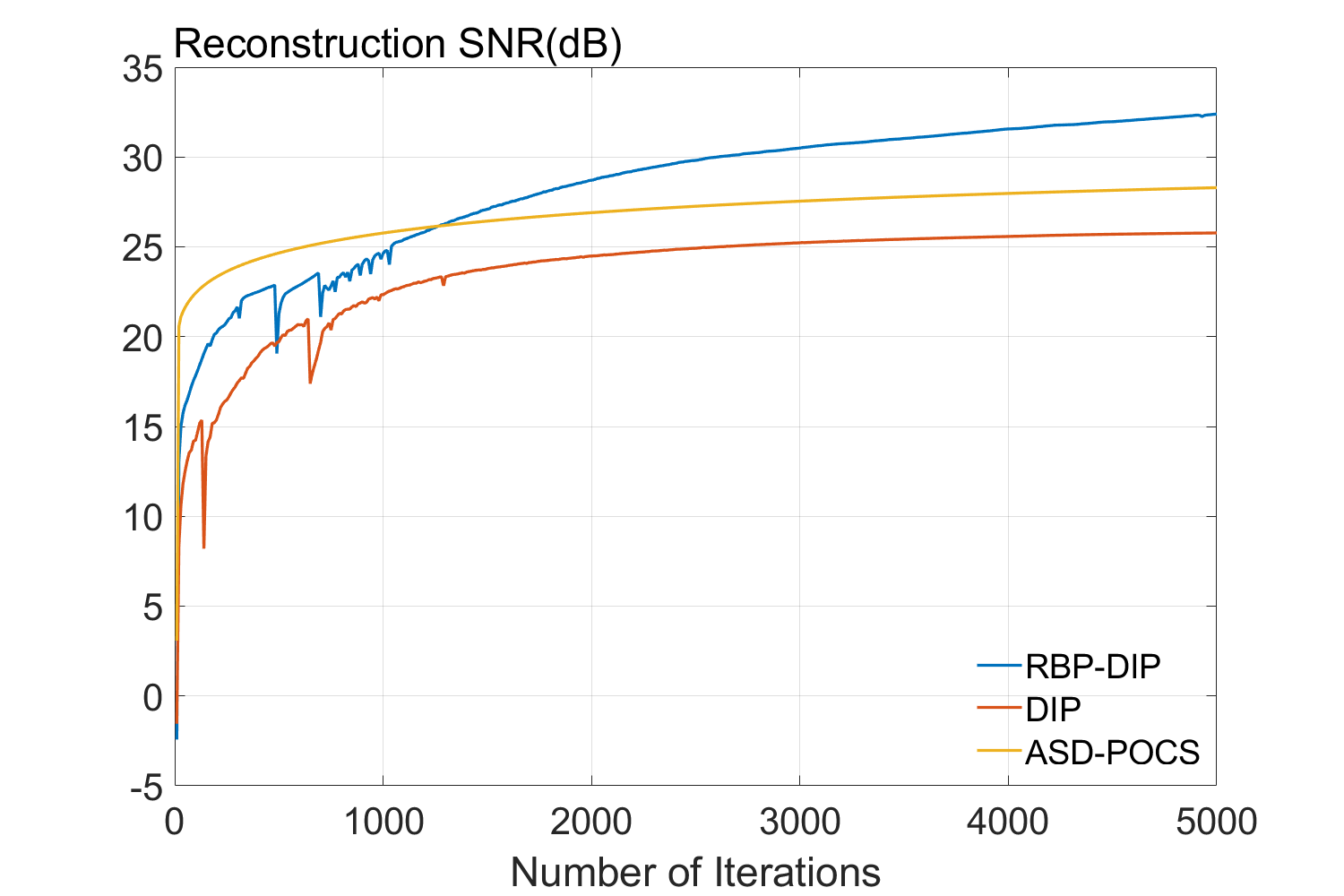}}
        \centerline{(b)}\medskip
    \end{minipage}
    \caption{The reconstruction loss ($l_2$ norm in logarithmic scale) and SNR (dB) of different methods under few-view condition (number of views is $90$, uniformly distributed from $0^\circ$ to $180^\circ$). On the one hand, RBP-DIP converges faster than DIP, demonstrating the effectiveness of our proposed RBP connection in improving the original DIP method. On the other hand, RBP-DIP achieves higher SNR with a loss larger than the ASD-POCS method. This demonstrates the superiority of the RBP-DIP algorithm, as such improvement is not merely caused by the further minimization of the loss function, but rather from the utilization of a better prior (the successful combination of conventional IR and deep image prior).}
    \label{trainingperformance}
\end{figure}

\section{Experiments and Results}\label{sec:exp}
In this section, we first conduct experiments showing the effectiveness of the proposed RBP connection. We also compare the differences between the proposed framework and the well-established IR algorithm ASD-POCS~\cite{sidky2008image} under multiple conditions. In that case, the effect of employing an untrained neural network for iterative reconstruction can be assessed. In addition, the proposed framework undergoes comparative analysis against the well-known DIP model~\cite{baguer2020computed}, as well as two widely recognized pre-trained models MED50~\cite{jin2017deep} and RED-CNN~\cite{chen2017low}. The structural similarities among these models show the merits of the proposed method while eliminating model complexity bias. The aforementioned methods are first tested under parallel-beam geometry with the reconstruction resolution $512 \times 512$. The Gram filtering method proposed by Shu and Entezari~\cite{shu2022exact} is used to calculate the forward and back projection exactly and efficiently. Next, $512 \times 512$ reconstructions under fan-beam geometry are conducted, where the ASTRA toolbox's~\cite{van2015astra} line\_fanflat projector is used for forward and back projection. Images from LIDC-IDRI~\cite{armato2011lung} (The Lung Image Database Consortium image collection) dataset are used in our experiments. To address potential issues related to inverse crime, sinogram data $\boldsymbol g$ is acquired from upsampled ($1024 \times 1024$) ground truth images. Finally, we tested our method on the real CT data from Finnish Inverse Problems Society~\cite{hamalainen2015tomographic,bubba2017tomographic}. The signal-to-noise ratio (dB) and the structural similarity index (SSIM) are calculated to evaluate the accuracy of reconstructions. However, in this paper, we only plot the SNR curve since the trends of both metrics are highly consistent. 

It is worth mentioning that we have also tried other forward and back projection methods, including lookup table~\cite{ha2017look}, box spline~\cite{zhang2019box}, and separable footprint~\cite{long20103d}. The experiments show that the reconstruction accuracy of the RBP-DIP algorithm indeed varies depending on the accuracy of the chosen projection operator, but it is always significantly higher than that of conventional IR and DIP algorithms using the same projection operator.

The aforementioned neural networks are implemented in Python with the PyTorch library. The RMSProp algorithm serves as the optimizer, with a learning rate set to $1 \times 10^{-4}$, which decreases by a factor of $0.9$ every $250$ iterations. The hyperparameters utilized in the RBP-DIP framework are assigned the values $n_s = 250$, and $n_c = 10$. We have tried using methods such as uniform, normal, Xavier uniform, and Xavier normal distributions for network parameter initialization but found no significant differences. Therefore, the experiments in our paper all used PyTorch's default uniform initialization method.

It is worth mentioning that each pre-trained model is retrained for different settings such as the number of views, and projection angular range. Moreover, during the training procedure, test images are employed for the direct assessment of the pre-trained models, while validation images are not utilized in this context. Although this training strategy is normally unreasonable, the purpose of our experiment is not to train a stable pre-trained model. Instead, it is to show that in our experiments, the best performance a neural network can achieve on test images is still far below expectations.

\subsection{Effect of Residual Back Projection in Reconstruction}
To demonstrate the efficacy of the RBP connection and show the details of the RBP-DIP's reconstruction process, the procedure of a limited-angle CT reconstruction is shown in Fig.\ref{RBP recon process}. Its first row illustrates the results of reconstruction across different iterations, whereas the second row presents the respective inputs fed into the U-net, which are updated through the RBP connection. Here, the number of views is set to $90$, uniformly distributed from $0^{\circ}$ to $90^{\circ}$. This scenario presents a challenging limited-angle CT reconstruction problem.

In the first iteration, as depicted in Fig.\ref{RBP recon process}a, the input image undergoes an update via the RBP connection prior to being fed to the U-net. Thus, this input is the normalized first iteration output of the implemented IR algorithm (normalized back projection image in our case). The output appears completely randomized since the entire neural network is randomly initialized.

In the 10th and 20th iterations (Fig.\ref{RBP recon process}b and Fig.\ref{RBP recon process}c), the DIP property effectively expedites the recovery of the object over its support. Of note are lack of artifacts commonly caused by having missing views in the data. The input images highlight the region which can be relatively accurately reconstructed by conventional IR methods. This can be used to guide the model in the later iteration. In our experiment, the model capitalizes on the input images more when reconstructing the upper-left and lower-right segments of the image, while relying predominantly on the DIP property for the reconstruction of the upper-right and lower-left parts.

In the 2000th iteration, as depicted in Fig.\ref{RBP recon process}d, the reconstruction result becomes relatively artifact-free. At this stage, the network input primarily emphasizes the edges to help the method improve the supporting area. Moreover, the RBP connection can rectify artifacts specific to convolutional neural networks. Evidence of this can be observed in the second row of Fig.\ref{RBP recon process}c, Fig.\ref{RBP recon process}d, and Fig.\ref{RBP recon process}e, which display distinct horizontal and vertical patterns. These patterns are mainly caused by the convolution operation in the U-net. In other words, the DIP and RBP parts of the proposed framework are able to mutually rectify each other's errors. Consequently, a high-quality reconstruction result is attainable, as shown in Fig.\ref{RBP recon process}e.

\subsection{Effect of Residual Back Projection with Deep Image Prior}

A direct comparison among the proposed RBP-DIP, the widely used IR method ASD-POCS, and the current untrained method~\cite{baguer2020computed} is made to show the advantages of the RBP-DIP method. Given that the number of views is set to $90$, uniformly distributed from $0^{\circ}$ to $180^{\circ}$, the scenario presents a challenging few-view CT reconstruction problem. The reconstruction performance in terms of reconstruction loss and SNR is shown in Fig.\ref{trainingperformance}. From Fig.\ref{trainingperformance}a, it is evident that compared to DIP, RBP-DIP can better minimize the objective function. Although the ASD-POCS method can further minimize the objective function, Fig.\ref{trainingperformance}b shows that RBP-DIP achieves the highest SNR. This indicates that when the problem is highly ill-posed, where multiple solutions can minimize the objective function, the result chosen by RBP-DIP is better than others. Another thing worth mentioning is that the loss of RBP-DIP shown in Fig.\ref{trainingperformance}a is fluctuating, which is in line with our analysis in Section \ref{sec:rbpdip}. This implies that using the conventional IR method to fine-tune the output RBP-DIP can further improve its reconstruction results. 

For a more direct comparison, the three aforementioned methods are tested on the Forbild phantom under few-view ($60$ views which are evenly distributed from $0^\circ$ to $360^\circ$), and limited-angle ($120$ views which are evenly distributed from $0^\circ$ to $120^\circ$). From Fig.\ref{Fresult}, it is clear that both the DIP and RBP-DIP exhibit superior performance compared to ASD-POCS, particularly in the reconstruction of detailed structures. The magnified views of these structures are available in the bottom left and top right corners of these images. In contrast to DIP, RBP-DIP is better at mitigating some neural network-specific artifacts, with the corresponding magnified view presented in the bottom right corner of these images.

To better evaluate the effectiveness of the RBP-DIP framework in real-world scenarios, multiple experiments using real CT images are conducted in the following sections.

\subsection{Few-View CT Reconstruction}
\begin{figure}[htb]
    \begin{minipage}[b]{\linewidth}
        \centering
        \centerline{\includegraphics[width=\linewidth]{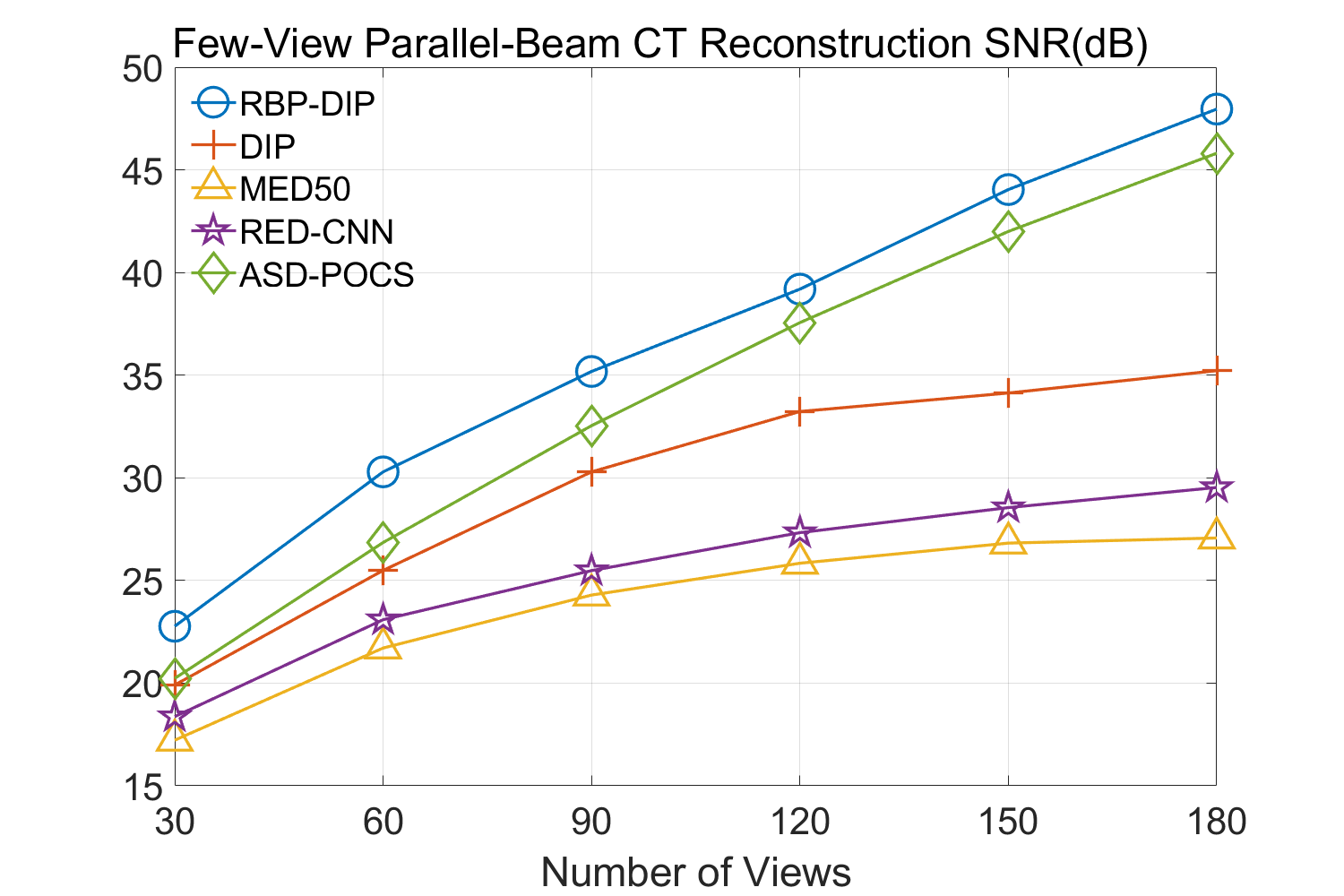}}
        \centerline{(a)}\medskip
    \end{minipage}
    
    \begin{minipage}[b]{\linewidth}
        \centering
        \centerline{\includegraphics[width=\linewidth]{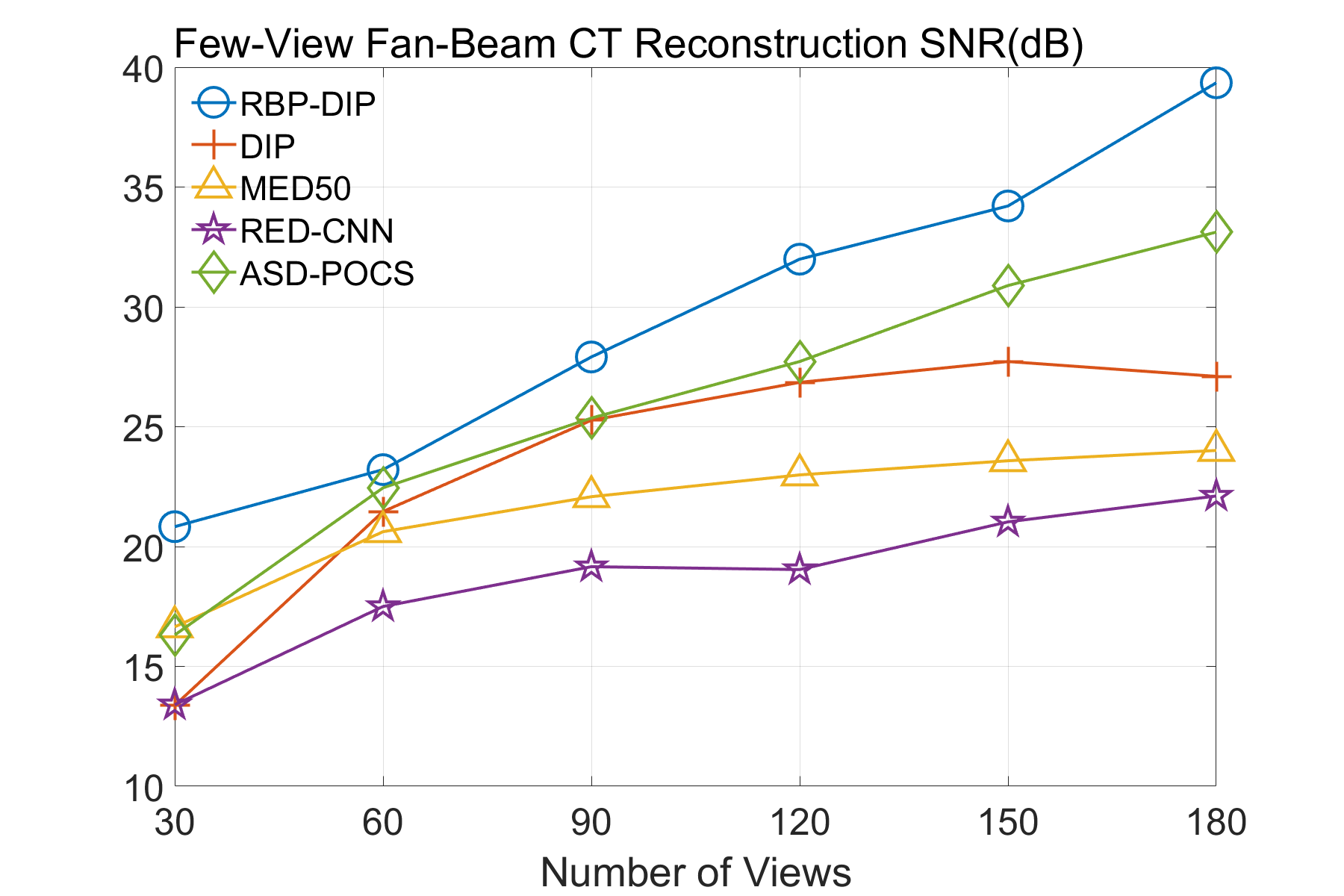}}
        \centerline{(b)}\medskip
    \end{minipage}
    \caption{The performance of few-view CT reconstruction for different methods under different numbers of views and geometries. The RBP-DIP method outperforms all other methods in all cases.}
    \label{SVRecon}
\end{figure}

In this section, the reconstruction performance of our proposed framework under few-view conditions will be tested. For the parallel and fan beam geometry, the number of views increases from $30$ to $180$, uniformly distributing from $0^\circ$ to $180^\circ$ and $0^\circ$ to $360^\circ$ respectively. Such settings provide a complete benchmark of reconstruction performance, ranging from extremely sparse to relatively complete, full-view CT reconstruction. The experiment results are shown in Fig.\ref{SVRecon}. Additionally, the ground truth, few-view ($30$ views), and full-view ($180$ views) CT reconstruction results of different methods are shown in the first and third rows of Fig.\ref{reconresult} (parallel-beam, LIDC-IDRI dataset), and Fig.\ref{reconresultfan} (fan-beam, LIDC-IDRI dataset).
 
\subsection{Limited-Angle CT Reconstruction}
\begin{figure}[htb]
    \begin{minipage}[b]{\linewidth}
        \centering
        \centerline{\includegraphics[width=\linewidth]{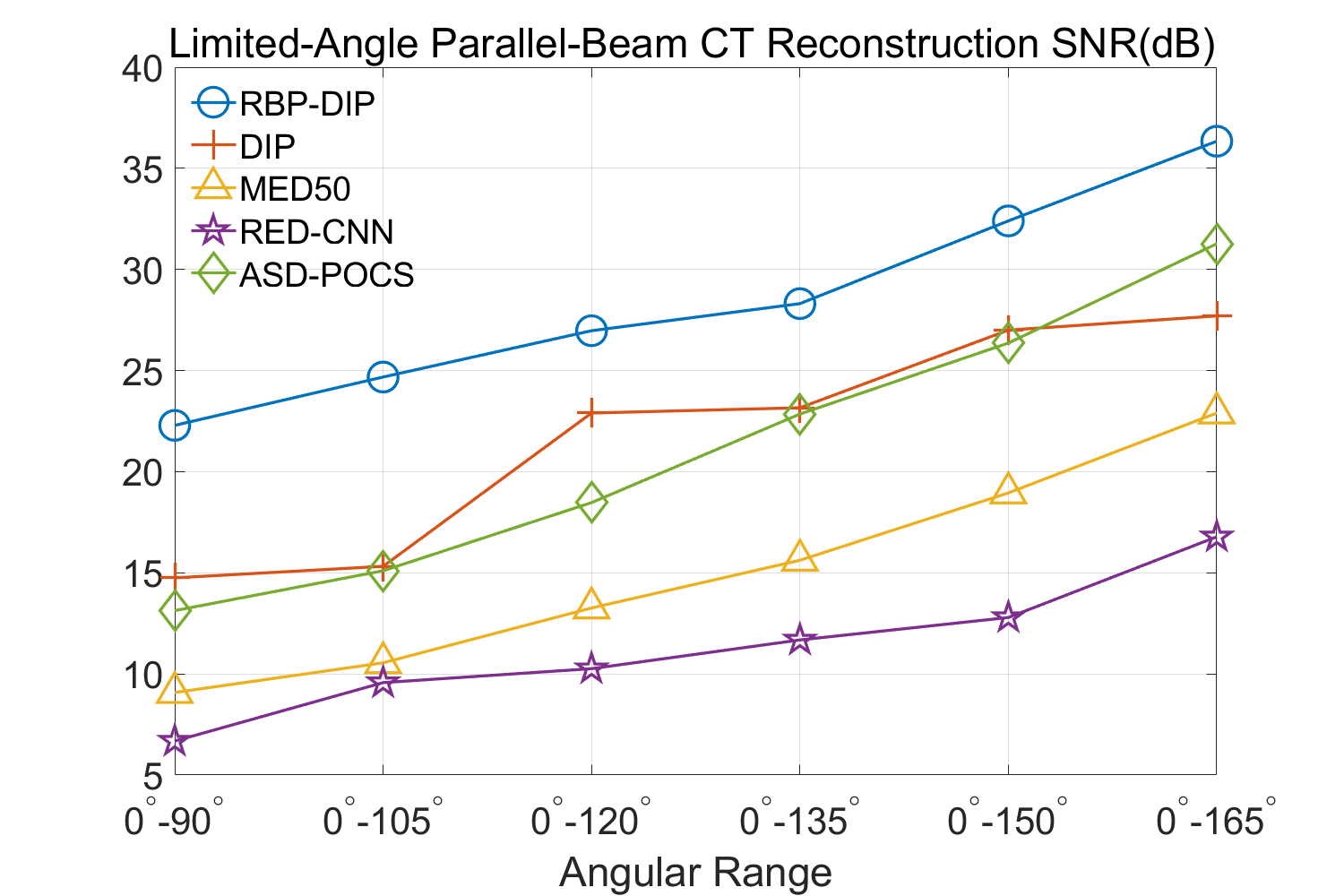}}
        \centerline{(a)}\medskip
    \end{minipage}
    
    \begin{minipage}[b]{\linewidth}
        \centering
        \centerline{\includegraphics[width=\linewidth]{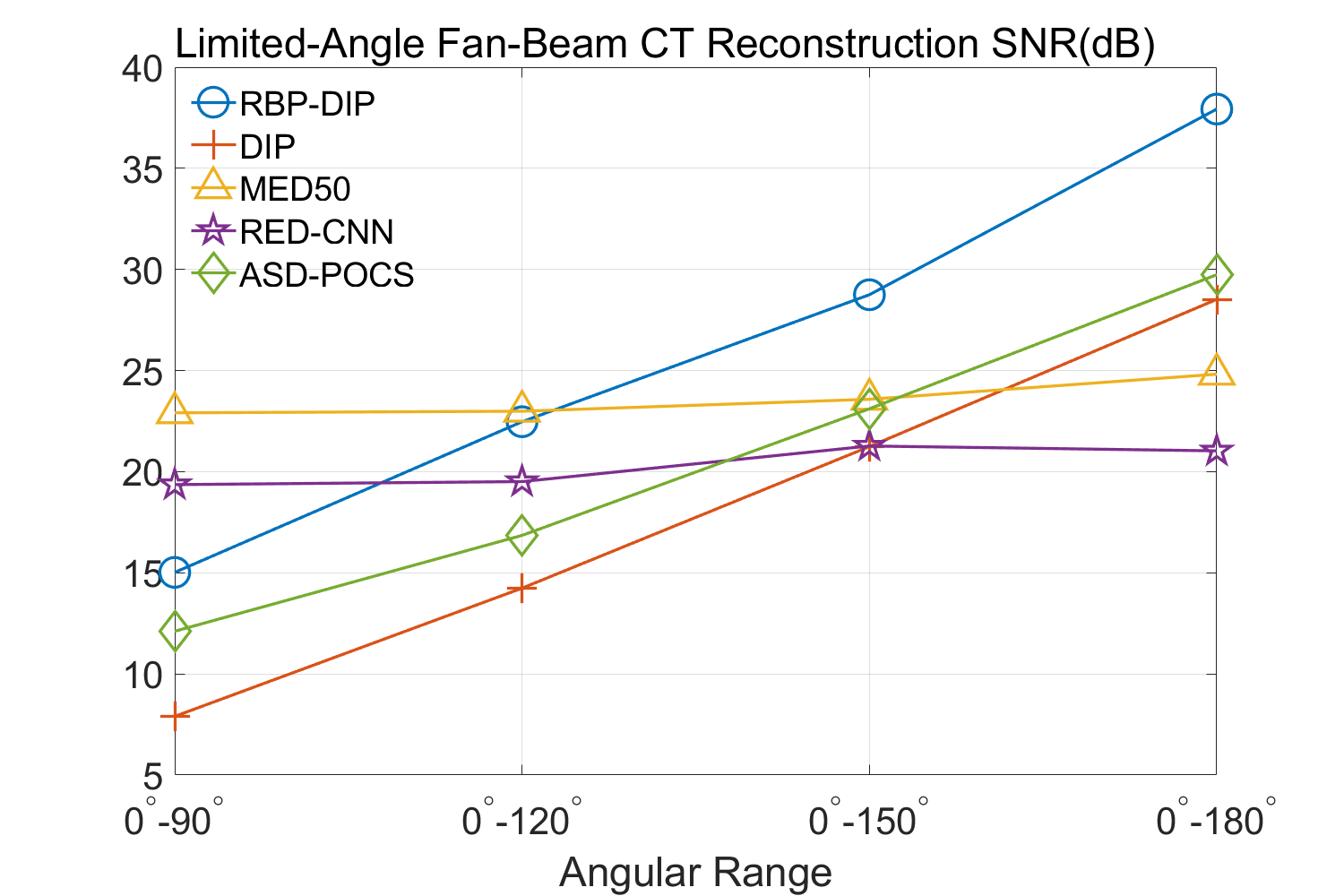}}
        \centerline{(b)}\medskip
    \end{minipage}
    \caption{The performance of limited-angle CT reconstruction for different methods under different angular ranges and geometries. The RBP-DIP method outperforms all other methods in most cases.}
    \label{LARecon}
\end{figure}

\begin{figure*}
	\centering
	\begin{minipage}[b]{.16\linewidth}
		\centering
		\centerline{\includegraphics[width=\linewidth]{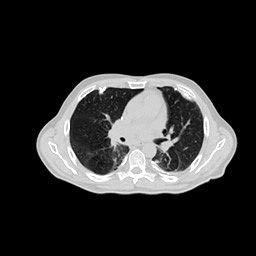}}
		\centerline{ }\medskip
	\end{minipage}
	\begin{minipage}[b]{.16\linewidth}
		\centering
        \centerline{SNR/SSIM}\medskip
		\centerline{\includegraphics[width=\linewidth]{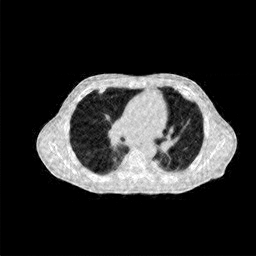}}
		\centerline{20.23dB/0.79}\medskip
	\end{minipage}
	\begin{minipage}[b]{.16\linewidth}
		\centering
        \centerline{SNR/SSIM}\medskip
		\centerline{\includegraphics[width=\linewidth]{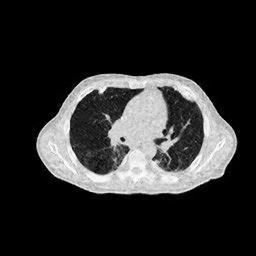}}
		\centerline{24.92dB/0.90}\medskip
	\end{minipage}
	\begin{minipage}[b]{.16\linewidth}
		\centering
        \centerline{SNR/SSIM}\medskip
		\centerline{\includegraphics[width=\linewidth]{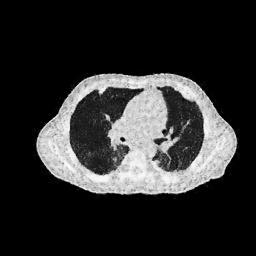}}
		\centerline{19.9dB/0.82}\medskip
	\end{minipage}
	\begin{minipage}[b]{.16\linewidth}
		\centering
        \centerline{SNR/SSIM}\medskip
		\centerline{\includegraphics[width=\linewidth]{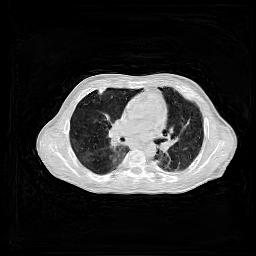}}
		\centerline{17.22dB/0.49}\medskip
	\end{minipage}
	\begin{minipage}[b]{.16\linewidth}
		\centering
        \centerline{SNR/SSIM}\medskip
		\centerline{\includegraphics[width=\linewidth]{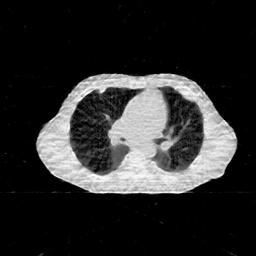}}
		\centerline{17.88dB/0.54}\medskip
	\end{minipage}

	\begin{minipage}[b]{.16\linewidth}
		\centering
		\centerline{\includegraphics[width=\linewidth]{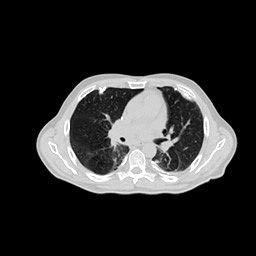}}
		\centerline{ }\medskip
	\end{minipage}
	\begin{minipage}[b]{.16\linewidth}
		\centering
		\centerline{\includegraphics[width=\linewidth]{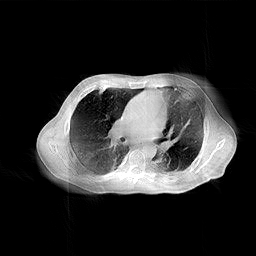}}
		\centerline{12.69dB/0.71}\medskip
	\end{minipage}
	\begin{minipage}[b]{.16\linewidth}
		\centering
		\centerline{\includegraphics[width=\linewidth]{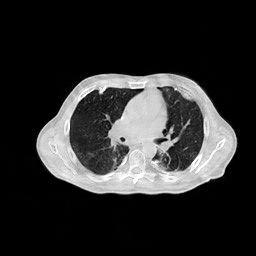}}
		\centerline{22.87dB/0.84}\medskip
	\end{minipage}
	\begin{minipage}[b]{.16\linewidth}
		\centering
		\centerline{\includegraphics[width=\linewidth]{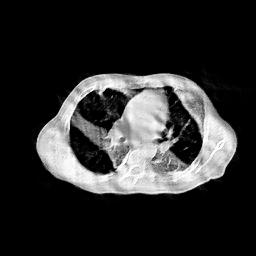}}
		\centerline{11.75dB/0.76}\medskip
	\end{minipage}
	\begin{minipage}[b]{.16\linewidth}
		\centering
		\centerline{\includegraphics[width=\linewidth]{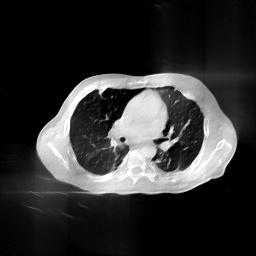}}
		\centerline{9.08dB/0.49}\medskip
	\end{minipage}
	\begin{minipage}[b]{.16\linewidth}
		\centering
		\centerline{\includegraphics[width=\linewidth]{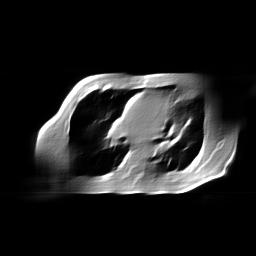}}
		\centerline{6.18dB/0.53}\medskip
	\end{minipage}
	\begin{minipage}[b]{.16\linewidth}
		\centering
		\centerline{\includegraphics[width=\linewidth]{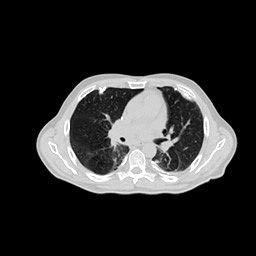}}
		\centerline{ }\medskip
		\centerline{(a) Ground Truth}\medskip
	\end{minipage}
	\begin{minipage}[b]{.16\linewidth}
		\centering
		\centerline{\includegraphics[width=\linewidth]{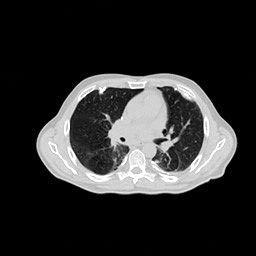}}
		\centerline{43.87dB/0.96}\medskip
		\centerline{(b) ASD-POCS}\medskip
	\end{minipage}
	\begin{minipage}[b]{.16\linewidth}
		\centering
		\centerline{\includegraphics[width=\linewidth]{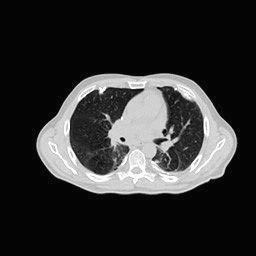}}
		\centerline{47.96dB/0.97}\medskip
		\centerline{(c) RBP-DIP}\medskip
	\end{minipage}
	\begin{minipage}[b]{.16\linewidth}
		\centering
		\centerline{\includegraphics[width=\linewidth]{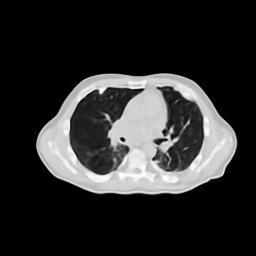}}
		\centerline{34.46dB/0.84}\medskip
		\centerline{(d) DIP}\medskip
	\end{minipage}
	\begin{minipage}[b]{.16\linewidth}
		\centering
		\centerline{\includegraphics[width=\linewidth]{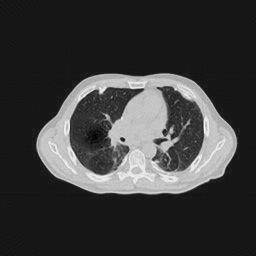}}
		\centerline{26.63dB/0.93}\medskip
		\centerline{(e) MED50}\medskip
	\end{minipage}
	\begin{minipage}[b]{.16\linewidth}
		\centering
		\centerline{\includegraphics[width=\linewidth]{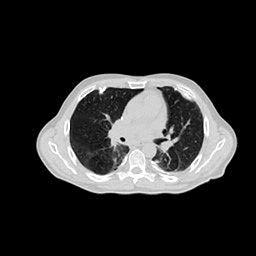}}
		\centerline{29.03dB/0.95}\medskip
		\centerline{(f) RED-CNN}\medskip
	\end{minipage}
	
	\caption{The parallel-beam reconstruction results of LIDC-IDRI dataset for different methods under few-view (first row, 30 projections distributed uniformly from $0^{\circ}$ to $180^{\circ}$), limited-angle (second row, 90 projections distributed uniformly from $0^{\circ}$ to $90^{\circ}$) and full-view (third row, 180 projections distributed uniformly from $0^{\circ}$ to $180^{\circ}$) conditions. The RBP-DIP method outperforms all other methods in all cases.}
	\label{reconresult}
\end{figure*}

\begin{figure*}
	\centering
	\begin{minipage}[b]{.16\linewidth}
		\centering
		\centerline{\includegraphics[width=\linewidth]{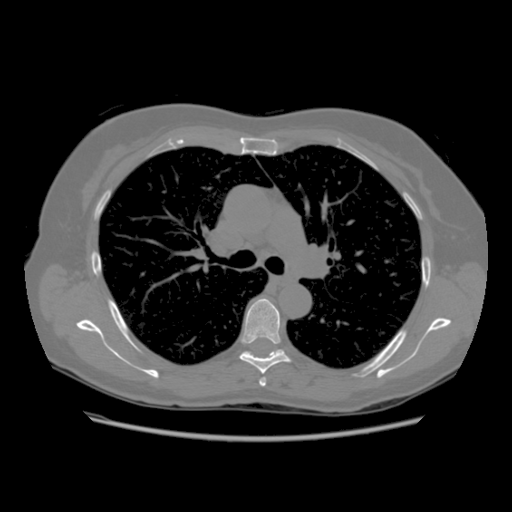}}
		\centerline{ }\medskip
	\end{minipage}
	\begin{minipage}[b]{.16\linewidth}
		\centering
        \centerline{SNR/SSIM}\medskip
		\centerline{\includegraphics[width=\linewidth]{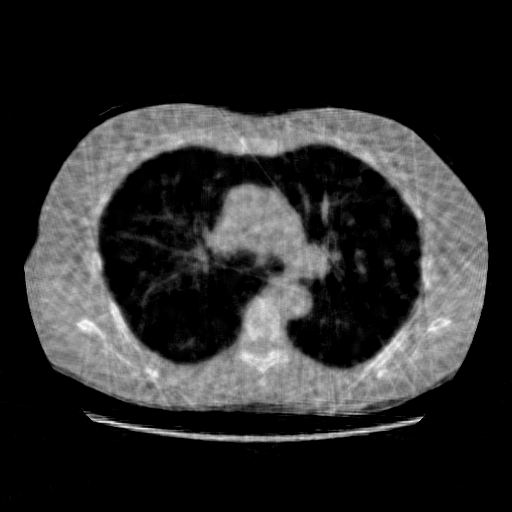}}
		\centerline{16.31dB/0.55}\medskip
	\end{minipage}
	\begin{minipage}[b]{.16\linewidth}
		\centering
        \centerline{SNR/SSIM}\medskip
		\centerline{\includegraphics[width=\linewidth]{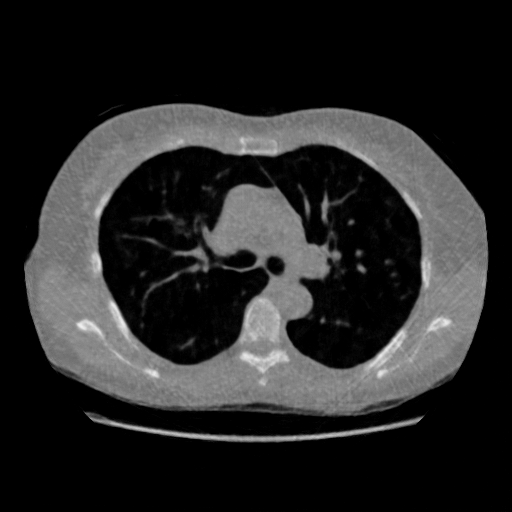}}
		\centerline{20.83dB/0.83}\medskip
	\end{minipage}
	\begin{minipage}[b]{.16\linewidth}
		\centering
        \centerline{SNR/SSIM}\medskip
		\centerline{\includegraphics[width=\linewidth]{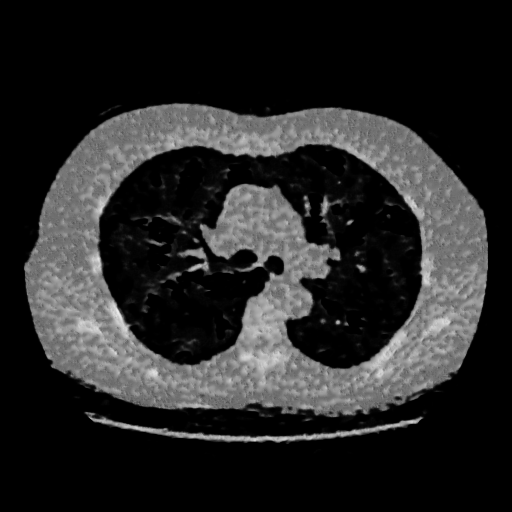}}
		\centerline{13.37dB/0.61}\medskip
	\end{minipage}
	\begin{minipage}[b]{.16\linewidth}
		\centering
        \centerline{SNR/SSIM}\medskip
		\centerline{\includegraphics[width=\linewidth]{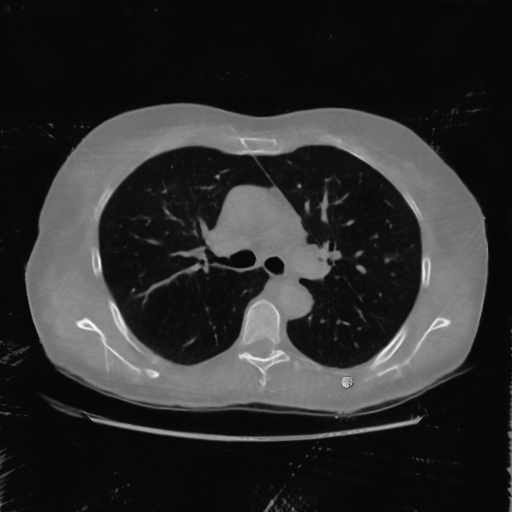}}
		\centerline{16.65dB/0.79}\medskip
	\end{minipage}
	\begin{minipage}[b]{.16\linewidth}
		\centering
        \centerline{SNR/SSIM}\medskip
		\centerline{\includegraphics[width=\linewidth]{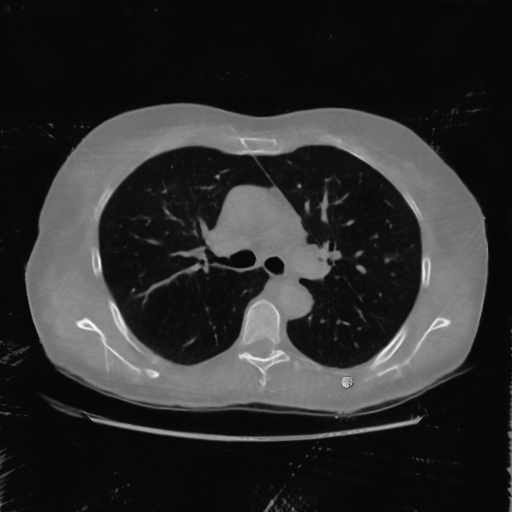}}
		\centerline{13.40dB/0.73}\medskip
	\end{minipage}

	\begin{minipage}[b]{.16\linewidth}
		\centering
		\centerline{\includegraphics[width=\linewidth]{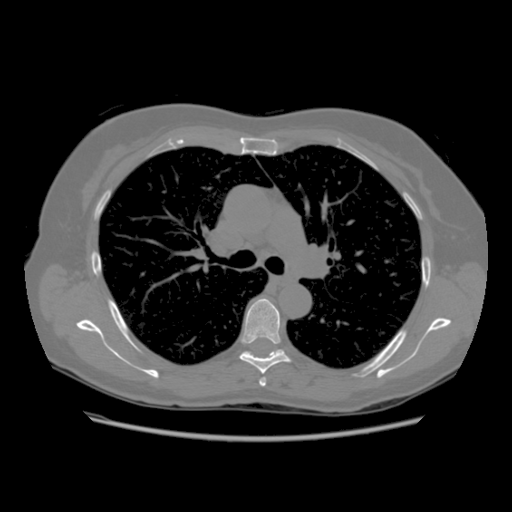}}
		\centerline{ }\medskip
	\end{minipage}
	\begin{minipage}[b]{.16\linewidth}
		\centering
		\centerline{\includegraphics[width=\linewidth]{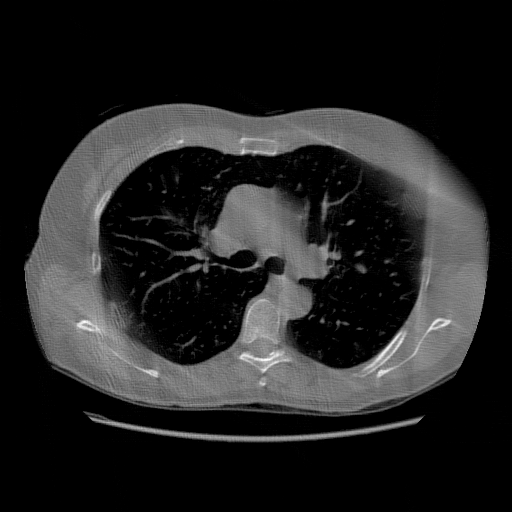}}
		\centerline{16.85dB/0.65}\medskip
	\end{minipage}
	\begin{minipage}[b]{.16\linewidth}
		\centering
		\centerline{\includegraphics[width=\linewidth]{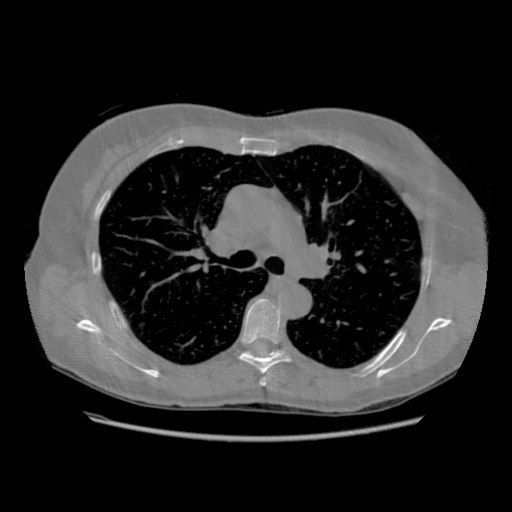}}
		\centerline{22.47dB/0.78}\medskip
	\end{minipage}
	\begin{minipage}[b]{.16\linewidth}
		\centering
		\centerline{\includegraphics[width=\linewidth]{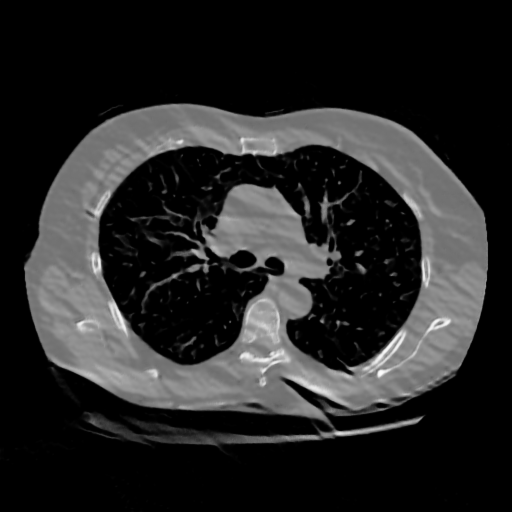}}
		\centerline{14.25dB/0.63}\medskip
	\end{minipage}
	\begin{minipage}[b]{.16\linewidth}
		\centering
		\centerline{\includegraphics[width=\linewidth]{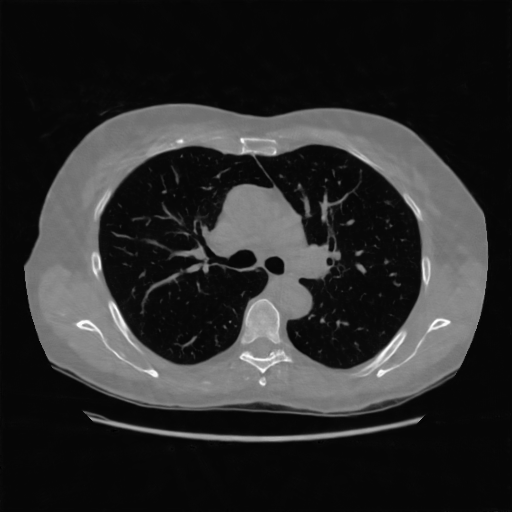}}
		\centerline{22.99dB/0.62}\medskip
	\end{minipage}
	\begin{minipage}[b]{.16\linewidth}
		\centering
		\centerline{\includegraphics[width=\linewidth]{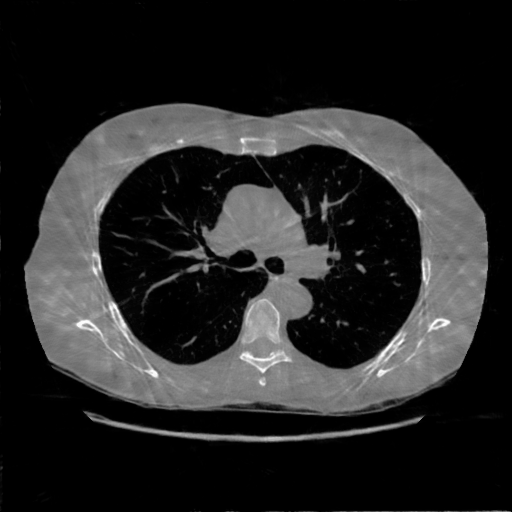}}
		\centerline{19.51dB/0.56}\medskip
	\end{minipage}
	\begin{minipage}[b]{.16\linewidth}
		\centering
		\centerline{\includegraphics[width=\linewidth]{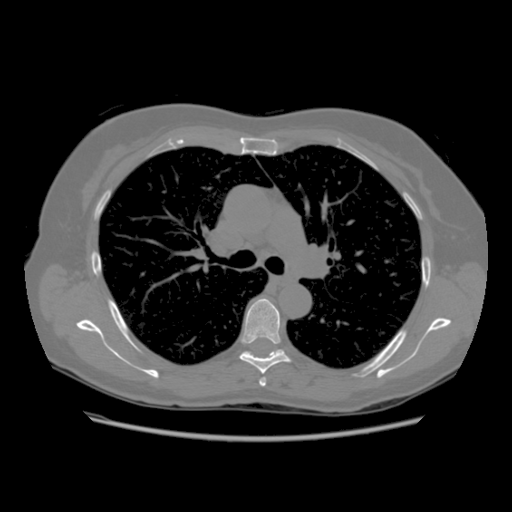}}
		\centerline{ }\medskip
		\centerline{(a) Ground Truth}\medskip
	\end{minipage}
	\begin{minipage}[b]{.16\linewidth}
		\centering
		\centerline{\includegraphics[width=\linewidth]{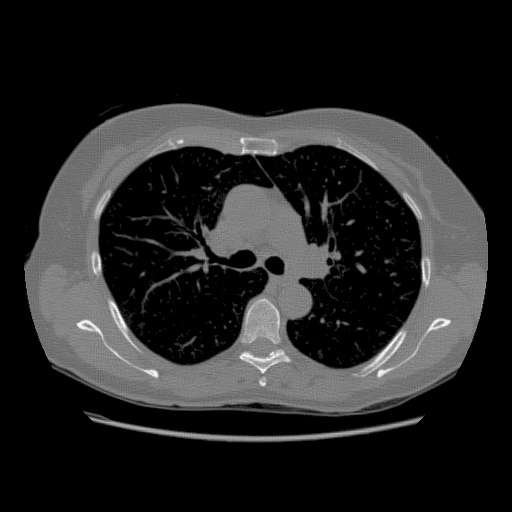}}
		\centerline{30.15dB/0.85}\medskip
		\centerline{(b) ASD-POCS}\medskip
	\end{minipage}
	\begin{minipage}[b]{.16\linewidth}
		\centering
		\centerline{\includegraphics[width=\linewidth]{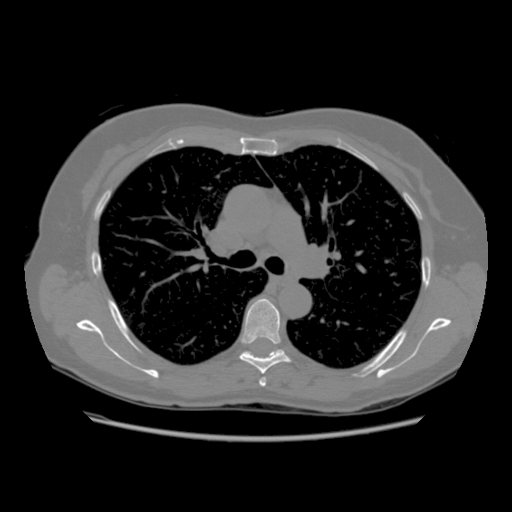}}
		\centerline{39.35dB/0.94}\medskip
		\centerline{(c) RBP-DIP}\medskip
	\end{minipage}
	\begin{minipage}[b]{.16\linewidth}
		\centering
		\centerline{\includegraphics[width=\linewidth]{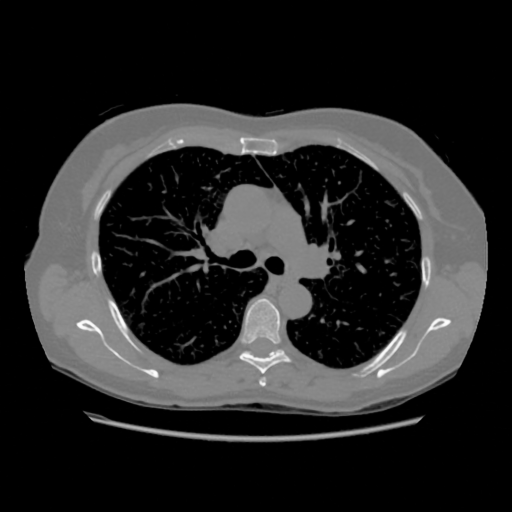}}
		\centerline{27.10dB/0.78}\medskip
		\centerline{(d) DIP}\medskip
	\end{minipage}
	\begin{minipage}[b]{.16\linewidth}
		\centering
		\centerline{\includegraphics[width=\linewidth]{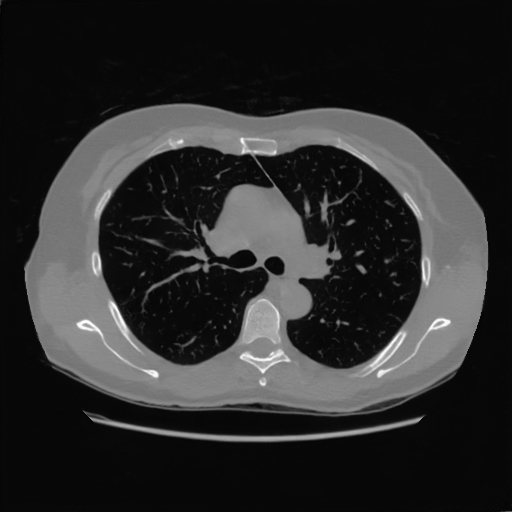}}
		\centerline{23.88dB/0.65}\medskip
		\centerline{(e) MED50}\medskip
	\end{minipage}
	\begin{minipage}[b]{.16\linewidth}
		\centering
		\centerline{\includegraphics[width=\linewidth]{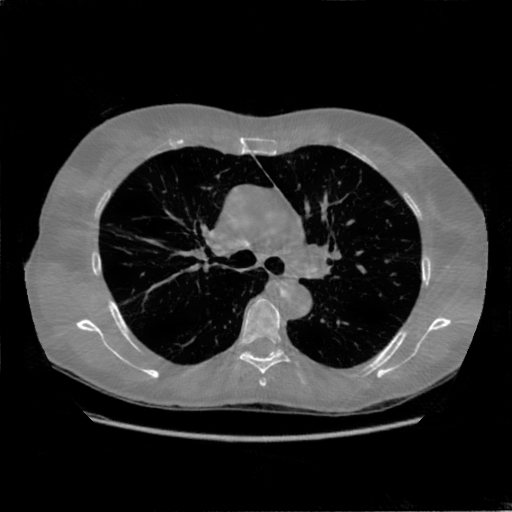}}
		\centerline{20.59dB/0.60}\medskip
		\centerline{(f) RED-CNN}\medskip
	\end{minipage}
	
	\caption{The fan-beam reconstruction results of LIDC-IDRI dataset for different methods under few-view (first row, 30 projections distributed uniformly from $0^{\circ}$ to $360^{\circ}$), limited-angle (second row, 120 projections distributed uniformly from $0^{\circ}$ to $120^{\circ}$) and dense-view (third row, 180 projections distributed uniformly from $0^{\circ}$ to $360^{\circ}$) conditions. The RBP-DIP method outperforms all other methods in most cases.}
	\label{reconresultfan}
\end{figure*}

To test the proposed framework’s performance on limited-angle reconstruction, we redo the experiment in the above section with the angular range changing from $0^\circ-90^\circ$ to $0^\circ-165^\circ$ for parallel-beam geometry and $0^\circ-90^\circ$ to $0^\circ-180^\circ$ for fan-beam geometry, one projection per degree. The experiment results are shown in Fig.\ref{LARecon}. Also, the ground truth and the limited-angle CT reconstruction results of different methods are shown in the row of Fig.\ref{reconresult} (parallel-beam, LIDC-IDRI dataset), and Fig.\ref{reconresultfan} (fan-beam, LIDC-IDRI dataset).

\subsection{Non-Ideal Factors}
\label{sec:P&N}
As previously mentioned, one of the most significant advantages of our proposed method lies in the elimination of the training process. This advantage precludes any potential hindrances arising from an insufficient training procedure or inconsistencies between inference input and training dataset. To examine this assertion, experiments under multiple non-ideal conditions have been conducted.

\subsubsection{Rotation}
\begin{figure}[htb]
    \begin{minipage}[b]{\linewidth}
        \centering
        \centerline{\includegraphics[width=\linewidth]{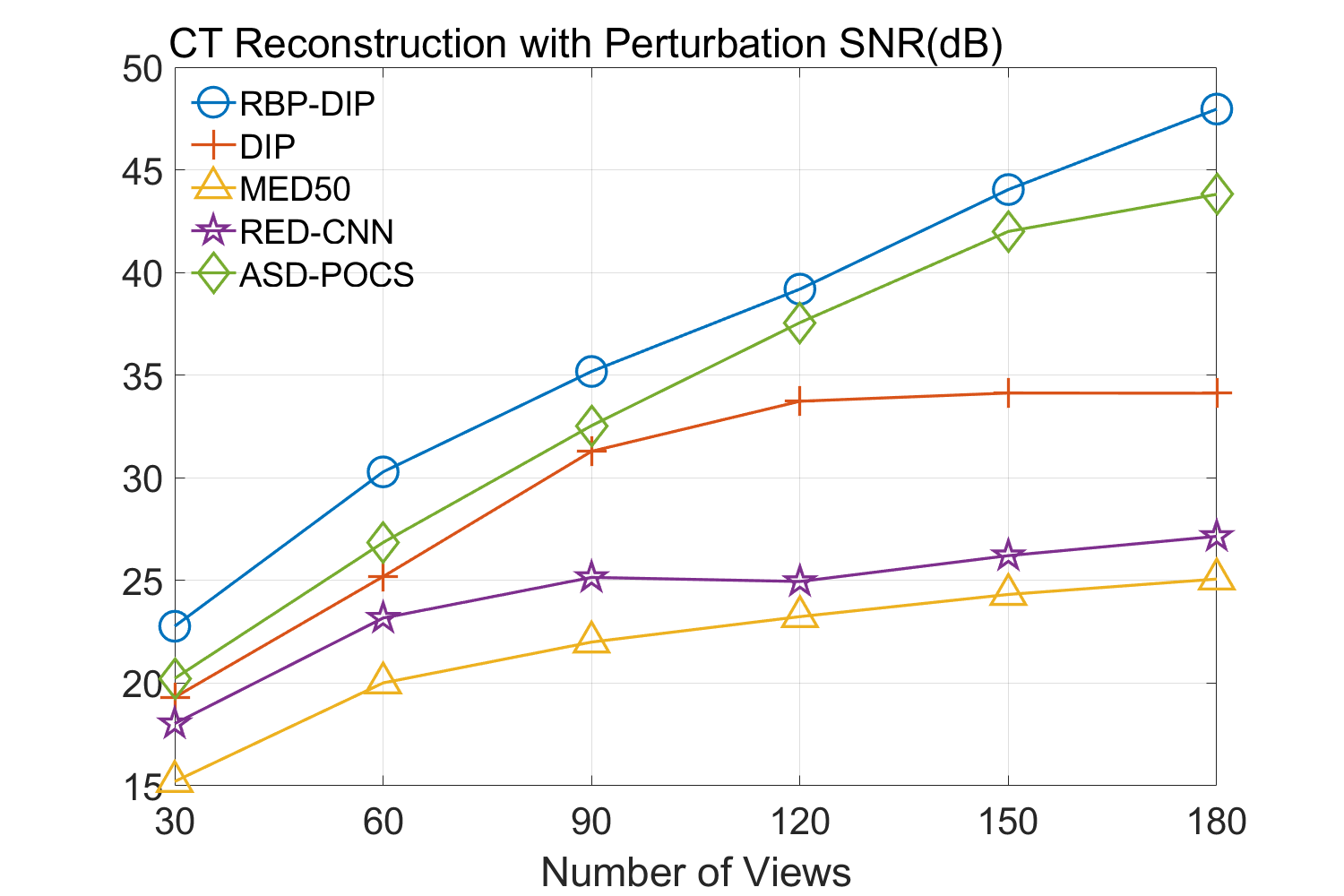}}
    \end{minipage}
    \caption{The performance of few-view CT reconstruction with perturbation (rotated by $30^\circ$) for different methods.}
    \label{SVreconR}
\end{figure}

\begin{figure}[htb]
    \centering
    \hfill
    \begin{minipage}[b]{.45\linewidth}
        \centering
        \centerline{\includegraphics[width=\linewidth]{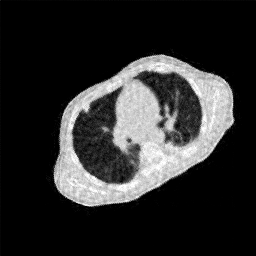}}
        \centerline{(a) ASD-POCS (19.92dB/0.76)}\medskip
    \end{minipage}
    \hfill
    \begin{minipage}[b]{.45\linewidth}
        \centering
        \centerline{\includegraphics[width=\linewidth]{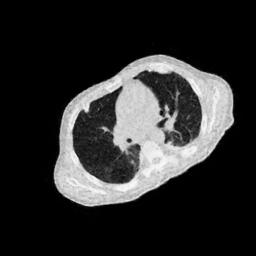}}
        \centerline{(b) RBP-DIP (23.82dB/0.88)}\medskip
    \end{minipage}
    
    \hfill
    \begin{minipage}[b]{.45\linewidth}
        \centering
        \centerline{\includegraphics[width=\linewidth]{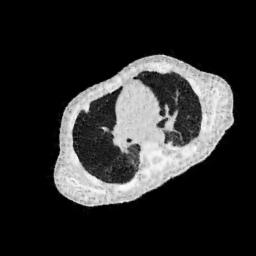}}
        \centerline{(c) DIP (21.88dB/0.80)}\medskip
    \end{minipage}
    \hfill
    \begin{minipage}[b]{.45\linewidth}
        \centering
        \centerline{\includegraphics[width=\linewidth]{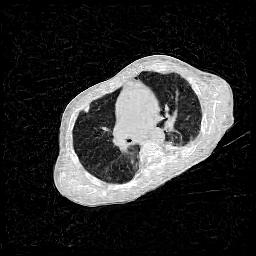}}
        \centerline{(d) MED50 (15.61dB/0.42)}\medskip
    \end{minipage}
    \caption{The reconstruction results and corresponding SNR/SSIM of the few-view ($30$ views) CT with perturbation (rotate by $30^\circ$) using different methods. The pre-trained model suffers from artifacts caused by the rotation.}
    \label{recon_rotate}
\end{figure}

The few-view CT reconstruction experiment is replicated, utilizing an identical test image rotated by $30^\circ$ degrees to simulate a rudimentary pose alteration of the patient. The reconstruction performances are shown in Fig.\ref{SVreconR}, and the reconstruction results corresponding to $30$ views are shown in Fig.\ref{recon_rotate}. A similar experiment is also conducted under the limited-angle condition, yielding similar results.

\subsubsection{Quantum Noise}
\begin{figure}[htb]
    \begin{minipage}[b]{\linewidth}
        \centering
        \centerline{\includegraphics[width=\linewidth]{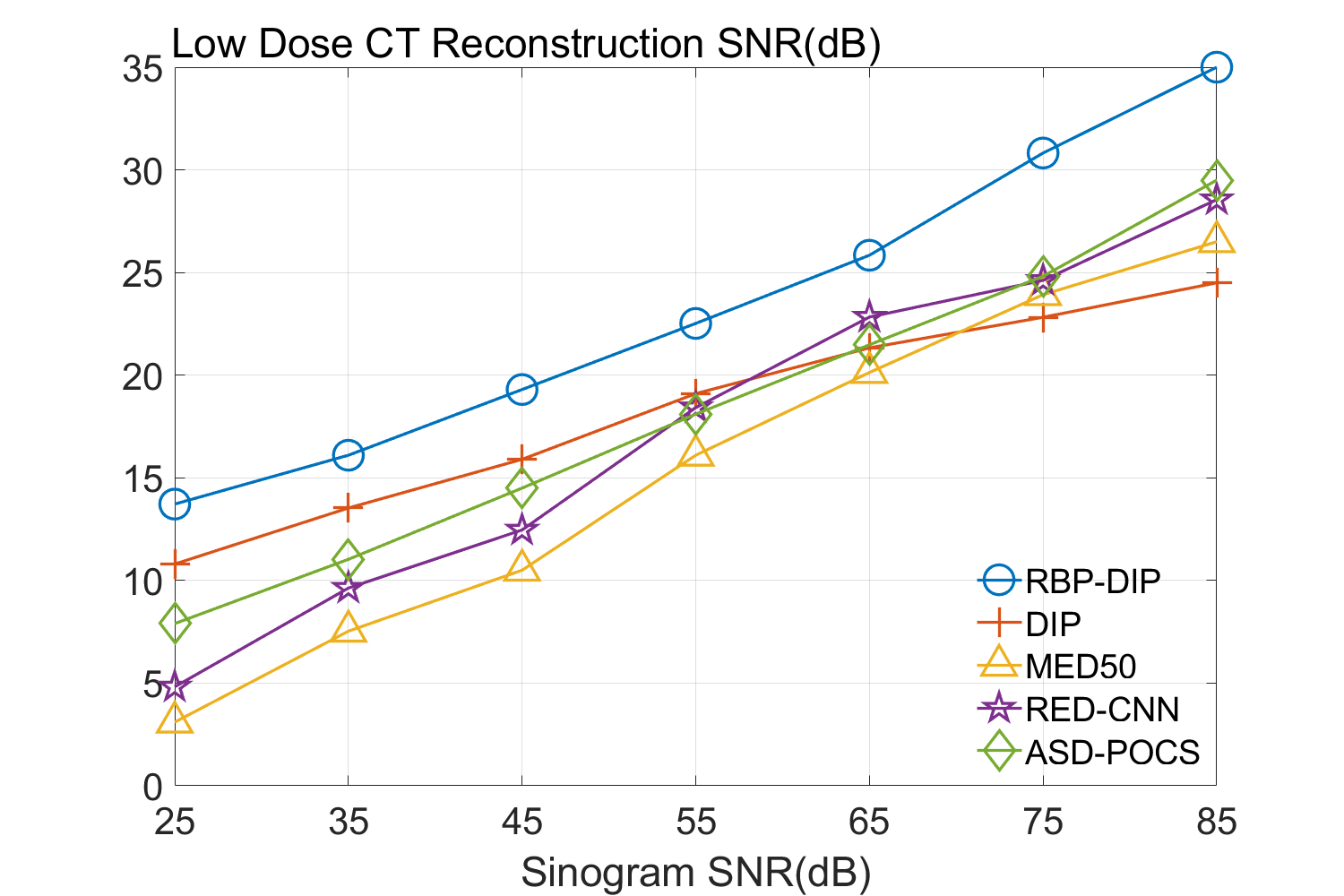}}
    \end{minipage}
    \caption{Impact of quantum noise on reconstruction SNR(dB). The RBP-DIP is the least susceptible to noise.}
    \label{lowdose}
\end{figure}

To evaluate the efficacy of the proposed algorithm under conditions of reduced dosage, the previously described reconstruction experiments are replicated utilizing sinogram data contaminated with noise following a Poisson distribution. The mean count of X-ray photons detected by the $i$th detector can be expressed as:
\begin{equation}\nonumber
    E_i = I_0e^{- \mu g_i}.
\end{equation}
In this experiment, a multiplication factor denoted by $\mu = 0.183/$cm was employed. This factor corresponds to an energy level of 80 keV, which approximates the mean energy of the beam utilized in a diagnostic CT scan conducted at a voltage of 120 kVp~\cite{yu2012simulation}. Throughout the experiment, $I_0$, the blank measurement, is assigned values ranging from $10^2$ to $10^8$, which correspond to sinogram SNR (dB) values of $25.6, 35.6, ..., 85.6$. It is worth mentioning that analogous experiments were conducted under both few-view and limited-angle scenarios, with the proposed method demonstrating a marked superiority over other methods across all conditions. This enhancement, however, is primarily attributed to the exceptional performance of the proposed method under sparse-measurement circumstances, rather than an increased resistance to noise. To facilitate an equitable comparison, the results displayed in Fig.\ref{lowdose} are solely derived from the full-view scenario, which encompasses an angular range of $0^\circ-180^\circ$ with one view per degree. It is worth mentioning that this experiment solely focuses on the resilience of the RBP-DIP framework itself against noise, thus not employing any existing methods such as total variation regularization and DeepRED~\cite{mataev2019deepred} for noise mitigation. However, these methods can be easily integrated into the RBP-DIP framework, as it shares a similar workflow with DIP.

\subsection{Real CT Reconstruction}
\begin{figure*}
	\centering
	\begin{minipage}[b]{.24\linewidth}
		\centering
		\centerline{\includegraphics[width=\linewidth]{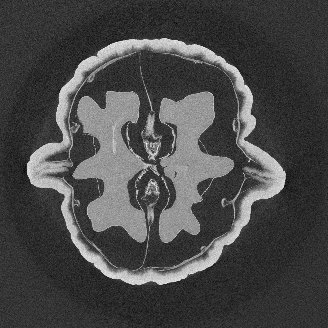}}
		\centerline{ }\medskip
	\end{minipage}
	\begin{minipage}[b]{.24\linewidth}
		\centering
        \centerline{SNR/SSIM}\medskip
		\centerline{\includegraphics[width=\linewidth]{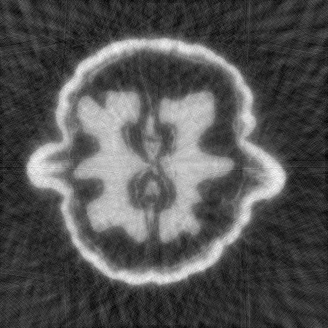}}
		\centerline{10.94dB/0.41}\medskip
	\end{minipage}
	\begin{minipage}[b]{.24\linewidth}
		\centering
        \centerline{SNR/SSIM}\medskip
		\centerline{\includegraphics[width=\linewidth]{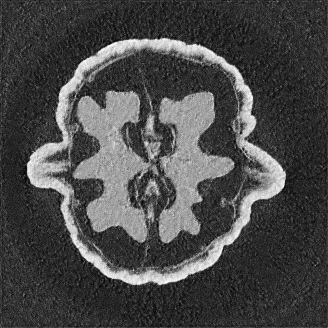}}
		\centerline{13.60dB/0.40}\medskip
	\end{minipage}
	\begin{minipage}[b]{.24\linewidth}
		\centering
        \centerline{SNR/SSIM}\medskip
		\centerline{\includegraphics[width=\linewidth]{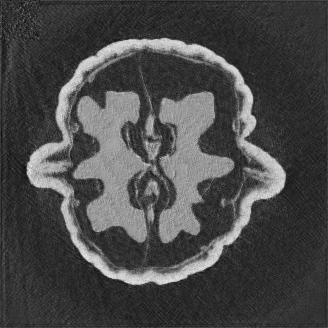}}
		\centerline{16.25dB/0.53}\medskip
	\end{minipage}

	\begin{minipage}[b]{.24\linewidth}
		\centering
		\centerline{\includegraphics[width=\linewidth]{1-gt.png}}
		\centerline{ }\medskip
	\end{minipage}
	\begin{minipage}[b]{.24\linewidth}
		\centering
		\centerline{\includegraphics[width=\linewidth]{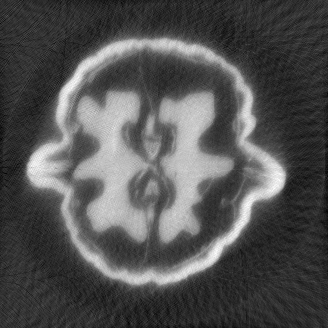}}
		\centerline{10.79dB/0.45}\medskip
	\end{minipage}
	\begin{minipage}[b]{.24\linewidth}
		\centering
		\centerline{\includegraphics[width=\linewidth]{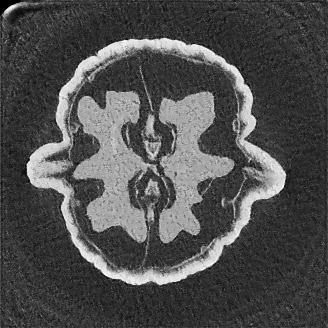}}
		\centerline{13.91dB/0.47}\medskip
	\end{minipage}
	\begin{minipage}[b]{.24\linewidth}
		\centering
		\centerline{\includegraphics[width=\linewidth]{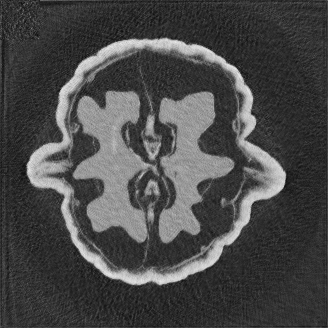}}
		\centerline{16.49dB/0.57}\medskip
	\end{minipage}
 	
	\begin{minipage}[b]{.24\linewidth}
		\centering
		\centerline{\includegraphics[width=\linewidth]{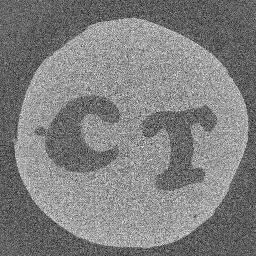}}
		\centerline{ }\medskip
	\end{minipage}
	\begin{minipage}[b]{.24\linewidth}
		\centering
		\centerline{\includegraphics[width=\linewidth]{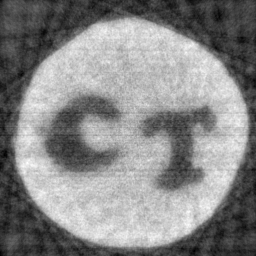}}
		\centerline{9.74dB/0.30}\medskip
	\end{minipage}
	\begin{minipage}[b]{.24\linewidth}
		\centering
		\centerline{\includegraphics[width=\linewidth]{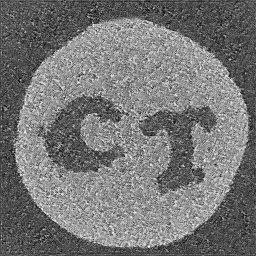}}
		\centerline{12.22dB/0.26}\medskip
	\end{minipage}
	\begin{minipage}[b]{.24\linewidth}
		\centering
		\centerline{\includegraphics[width=\linewidth]{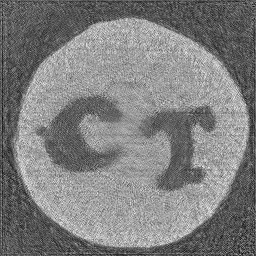}}
		\centerline{14.46dB/0.42}\medskip
	\end{minipage}
	
	\begin{minipage}[b]{.24\linewidth}
		\centering
		\centerline{\includegraphics[width=\linewidth]{2-gt.png}}
		\centerline{ }\medskip
		\centerline{(a) Reference}\medskip
	\end{minipage}
	\begin{minipage}[b]{.24\linewidth}
		\centering
		\centerline{\includegraphics[width=\linewidth]{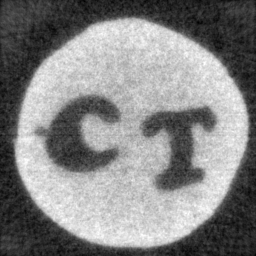}}
		\centerline{8.41dB/0.34}\medskip
		\centerline{(b) ASD-POCS}\medskip
	\end{minipage}
	\begin{minipage}[b]{.24\linewidth}
		\centering
		\centerline{\includegraphics[width=\linewidth]{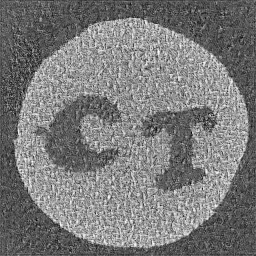}}
		\centerline{12.77dB/0.31}\medskip
		\centerline{(c) DIP}\medskip
	\end{minipage}
	\begin{minipage}[b]{.24\linewidth}
		\centering
		\centerline{\includegraphics[width=\linewidth]{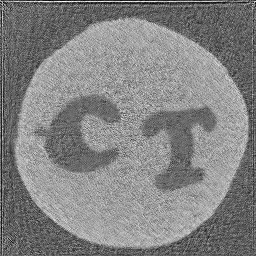}}
		\centerline{14.59dB/0.45}\medskip
		\centerline{(d) RBP-DIP}\medskip
	\end{minipage}
	\caption{The fan-beam reconstruction results of a walnut and a carved cheese. The data of the reference images is from the Finnish Inverse Problems Society. The first and third rows correspond to few-view CT reconstructions (30 projections distributed uniformly from $0^{\circ}$ to $360^{\circ}$). The second and fourth rows correspond to limited-view CT reconstructions (60 projections distributed uniformly from $0^{\circ}$ to $120^{\circ}$). Each column shows the reconstruction results of different methods. The first column shows the reference images which are generated by the FBP algorithm using full-view and high-resolution data.}
	\label{reconresult3}
\end{figure*}

To further verify the effectiveness of the RBP-DIP framework, real CT data from the Finnish Inverse Problem Society~\cite{hamalainen2015tomographic,bubba2017tomographic} was used to validate our algorithm. The few-view and limited-angle reconstruction results are shown in Figure \ref{reconresult3}. Due to the lack of available training sets, pre-trained models were not included in the experiments. Different from the previous experiment, data from Finnish Inverse Problem Society is highly noisy. However, it is still evident that the RBP-DIP outperforms all other algorithms in all experiments, even under high noise and highly ill-posed conditions.

\section{Discussion}\label{sec:diss}
\subsection{Comparisons on DIP, IR, and RBP-DIP}
It is worth mentioning that both DIP and RBP-DIP are iterative reconstruction methods since all these methods reconstruct images by minimizing an objective function on inference data iteratively. This minimization process offers a dependable lower bound, ensuring that their performance does not fall below that of conventional IR methods if both obtain similar loss. The distinction between conventional IR and DIP related methods lies in the use of an untrained convolutional neural network.

Conventional IR methods, even those with the help of regularizations such as total variation, are prone to artifacts when constrained by few-view and limited-angle conditions. However, as illustrated in the first and second rows of Fig.\ref{reconresult}b, and Fig.\ref{reconresultfan}b, these images still contain meaningful information which can be used to guide DIP related reconstruction methods, despite the presence of artifacts.

DIP related methods, which leverage the hierarchical structure of neural networks as a powerful prior, can better handle the aforementioned challenge. However, the original DIP method has its own limitations. It cannot generate detailed images or effectively enhance its accuracy as the number of measurements increases. For instance, in Fig.\ref{SVRecon}a, the ASD-POCS algorithm achieves an approximate SNR  gain of $25$dB when the number of views increases from $30$ to $180$, while the DIP method only attains an approximate $15$dB gain. This problem is also shown in the last row of Fig.\ref{reconresult}, and Fig.\ref{reconresultfan}. Moreover, the DIP method may produce neural network specific artifacts, as shown in Fig.\ref{reconresult}d, Fig.\ref{reconresultfan}d, and Fig.\ref{reconresult3}c. These artifacts are particularly problematic as they are often considered more undesirable than streak artifacts. Radiologists, with their professional experience, can interpret and account for streak artifacts, whereas network specific artifacts may prove more challenging to identify and address.

The proposed RBP-DIP framework combines the two approaches utilizing the newly devised RBP connection so that inherits the benefits of both methods. In Fig.\ref{trainingperformance}, the RBP-DIP method's attainment of a 5dB SNR enhancement over the ASD-POCS method, despite exhibiting a larger loss. \red{This implies that the improvement is not caused by the further minimization of the loss function but by the ability to choose a more reasonable reconstruction result from all candidates with similar losses.} This is precisely the direct evidence of the successful combination of conventional IR and deep image prior in the proposed RBP-DIP framework. Subsequently, the improvement surpassing the original DIP method indicates the efficacy of the RBP connection. Moreover, by employing the RBP connection, neural network specific artifacts can be rectified effectively. As a result, substantial advancements can be shown in Fig.\ref{SVRecon}, Fig.\ref{LARecon}, Fig.\ref{reconresult}, and Fig.\ref{reconresultfan}.

\subsection{Comparing with Pre-Trained Models}
We have also conducted comparative analyses with pre-trained models MED50 and RED-CNN under multiple conditions. These models share similar network structures and complexities with the RBP-DIP framework. We have opted not to examine models with more complicated networks, as the proposed method is untrained, necessitating a focus on the structural properties of the networks for equitable comparisons. Theoretically, it is expected that pre-trained models should obtain better reconstruction results than that of the proposed method, especially when the number of measurements is not enough, as they can acquire extra information from training datasets. However, as shown in Fig.\ref{SVRecon} and Fig.\ref{LARecon}, RBP-DIP outperforms the two pre-trained models in most cases, with the only exception being limited-angle fan-beam reconstruction where the angular range is $0^\circ-90^\circ$.

\begin{figure}[htb]
    \centering
    \hfill
    \begin{minipage}[b]{.45\linewidth}
        \centering
        \centerline{\includegraphics[width=\linewidth]{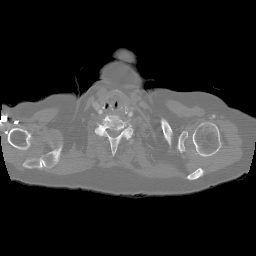}}
        
    \end{minipage}
    \hfill
    \begin{minipage}[b]{.45\linewidth}
        \centering
        \centerline{\includegraphics[width=\linewidth]{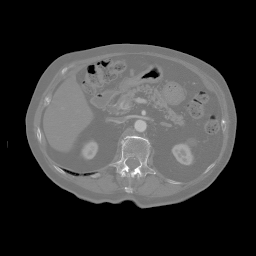}}
    \end{minipage}
    \medskip
    \caption{Training instances in LIDC-IDRI dataset. Some of the training instances may even downgrade the performance of pre-trained models due to the inconsistency in postures and CT slices.}
    \label{badtraining}
\end{figure}

Pre-trained models' low performances may be caused by the limited number of views. In the original research, the network inputs are high-quality FBP images (with 400 views), which possess sufficient projections and relatively minimal artifacts. In that case, the inconsistencies between the input and the corresponding output are relatively small, so that pre-trained models can learn the correct prior with limited training data. In our experiments, the FBP images always suffer from severe artifacts, and the aforementioned differences are relatively large. As a result, pre-trained models cannot be trained properly without a large, high-quality training dataset. In fact, some training instances in the current training dataset may even downgrade the performance of pre-trained models due to discrepancies in postures and CT slices, as illustrated in Fig.\ref{badtraining}.

Also, Fig.\ref{SVRecon} and Fig.\ref{LARecon}, particularly Fig.\ref{LARecon}b, show that an increasing number of views does not sufficiently lead to improved performance of pre-trained models. The reason is that pre-trained models aim to learn the mapping between the input and the corresponding output from training datasets, rather than actually solve the corresponding inverse problem. Increasing the number of views cannot directly strengthen this mapping. Conversely, the proposed RBP-DIP directly minimizes the inconsistency between the ground truth and reconstructed images under the same measurements. Increasing the number of views reduces the dimensions of the solution space and thus benefits both network optimization and the iterative reconstruction (IR) algorithms integrated in the RBP connection. This outcome is further verified in the last row of Fig.\ref{reconresult}, and Fig.\ref{reconresultfan}, where RBP-DIP considerably outperforms other methods.

\subsection{Non-Ideal Factors} 
In practical applications, the imaged object might exhibit inconsistencies with the training dataset due to various non-ideal factors. The $30^\circ$ rotation operation in Section \ref{sec:P&N} serves as a simplified simulation of such discrepancies, which may arise from differing poses of patients. The impact on reconstruction accuracy is shown in Fig.\ref{SVreconR}. The comparison between Fig.\ref{SVRecon}b and Fig.\ref{SVreconR} reveals that the pre-trained MED50 method is predominantly affected by a considerable decline in SNR. The reconstruction results are also presented in Fig.\ref{recon_rotate}. In contrast to the first row of Fig.\ref{reconresult}, the MED50 reconstruction result lacks a completely black background and exhibits horizontal artifacts in the non-empty region. This observation suggests that MED50 cannot effectively handle the perturbations in the inference data. A similar result is also obtained when adding quantum noise to the sinogram.

The proposed RBP-DIP framework does not necessitate any training images and attains the highest reconstruction accuracy in both experiments. Moreover, additional constraints or regularizations can be directly incorporated into the objective function or indirectly integrated into the RBP connection, enabling enhanced handling of various factors without retraining.

\subsection{Future Works and Prospects}
This study shows that the RBP-DIP framework shows improvements over the original DIP method and two pre-trained methods with similar network structures. Also, the RBP-DIP framework has great potential for further improvement. The RBP connection in this paper is implemented by a basic IR algorithm with the objective function $||\boldsymbol g - \boldsymbol A \boldsymbol x||^2$ without any additional constraints or regularizations. However, it is evident that RBP-DIP is compatible with all other IR methods, constraints, and regularizations. Notably, these techniques can be incorporated into both the objective function and the RBP connection. In contrast to pre-trained models, these adjustments can be adapted on demand (e.g., augmenting the total variation regularization's weight when dealing with noisy input data). Moreover, the proposed framework employs a basic U-net structure, but more sophisticated models such as U-net++\cite{zhou2018unet++} and ResUnet\cite{zhang2018road} might be integrated for further improvement.

As an enhancement of the DIP method, the challenges faced by the RBP-DIP algorithm are similar to the DIP:
\begin{itemize}
    \item Computational time: Unlike the inference process of pre-trained neural networks, where the network is initialized with learned weights, the RBP-DIP network starts from a random initialization, which implies that the proposed method is time-consuming. The reconstruction process of the RBP-DIP algorithm is the same as the training process of an ordinary neural network. The only difference is that the DIP and RBP-DIP algorithms need to be optimized on a single inference case, while ordinary neural networks need to be optimized on the entire training set. Considering the powerful parallel processing capabilities of current GPUs, the time to perform one forward and backward pass on a single case is almost the same as that for an entire batch of cases. In that case, the reconstruction speed of the RBP-DIP algorithm is $\frac{N}{M}$ times the training speed of a neural network with a training set size of $N$, a batch size of $M$, and the same network structure. This also matches our experimental results. In our future research, we will dedicate our efforts to exploring strategies such as employing a warm start and simplifying the network architecture to accelerate the RBP-DIP framework. Existing frameworks such as ADMM can also accelerate our method, which has been proven by researchers such as Mataev et al.~\cite{mataev2019deepred}.
    \item Noise sensitivity: Although superior to the original DIP algorithm, RBP-DIP remains relatively sensitive to noise. Thanks to the similarity in the workflows of the RBP-DIP and DIP, most noise mitigation methods~\cite{liu2019image,mataev2019deepred,jo2021rethinking} applicable to DIP are also suitable for RBP-DIP. Therefore, in future research, we will dedicate efforts to integrating the RBP-DIP algorithm with existing noise reduction techniques to further enhance its performance.
\end{itemize}

\section{Conclusion}\label{sec:conclusion}
In this paper, we present an untrained neural network related framework, RBP-DIP, for X-ray CT reconstruction. Remarkable reconstruction results are achieved using RBP-DIP without any training process or extra regularization, particularly under highly ill-posed conditions. The improvements are primarily attributable to the employment of the untrained neural network and the RBP connection, which utilize the advantages of DIP and conventional IR algorithms, respectively.

In comparison with conventional IR methods, RBP-DIP attains a higher reconstruction SNR, especially when the corresponding inverse problem is highly ill-posed. In comparison with the original DIP method, RBP-DIP exhibits superior accuracy and is free from neural network specific artifacts. These findings suggest that our proposed methodology successfully utilizes the advantages of both IR and DIP methods.

Comprehensive comparisons are also conducted with pre-trained models possessing similar network structures to RBP-DIP. The results indicate that obtaining a well-trained model necessitates a substantial, high-quality training dataset. The experiment subjected to perturbations reveals that a pre-trained network is biased toward training data. Such a problem is even more severe in the field of medical imaging, as deviation from normal is the key to diagnosis. In contrast, our untrained RBP-DIP leverages the hierarchical properties of convolutional neural networks without capitalizing on what the training data deems normal. These significant improvements further substantiate the superiority of RBP-DIP.

The proposed framework can handle multiple geometries and non-ideal factors. Such versatility exists thanks to the U-net architecture's widespread use in 2D and 3D images and the thorough analysis of IR algorithms in the RBP connection. In fact, the RBP-DIP framework is not limited to CT and has the potential to solve other reconstruction problems, as it only requires an untrained model for DIP image generation and a conventional IR method for residual back projection. Thanks to its concise architecture, further improvements such as employing more delicate IR methods and neural networks are also possible. Additional priors can also be incorporated as needed in both the RBP connection and the objective function.

In conclusion, the proposed framework further improves the original DIP algorithm for CT reconstruction, particularly under conditions of few views and limited angles.

\bibliographystyle{IEEEtran}
\bibliography{refs}

\end{document}

%% file: macros.tex
\usepackage{graphics,graphicx,url,color}
\usepackage{amsmath,amsfonts,amssymb,theorem,euscript,mathrsfs,bbm}
\newtheorem{theorem*}{Theorem}
\usepackage{epsf,epsfig,wrapfig}
\usepackage{array,enumitem,multicol}
\usepackage{footmisc}
\DeclareGraphicsExtensions{.ps}

\definecolor{navyblue}{rgb}{0.0, 0.0, 0.5}
\definecolor{lightblue}{rgb}{0.3, 0.3, 0.9}

\newcommand{\bracketprod}[2]{\left[{\kern-.75ex}\left[#1,#2\right]{\kern-.75ex}\right]_{}}

\DeclareMathOperator*{\argmin}{arg\,min}

\newcommand{\shah}{{\textstyle \amalg{\kern-4.pt\amalg}}}

